\documentclass[a4paper,twocolumn,11pt,accepted=2024-12-16]{quantumarticle}
\pdfoutput=1
\usepackage[utf8]{inputenc}
\usepackage[english]{babel}
\usepackage[T1]{fontenc}
\usepackage{amsmath}
\usepackage{hyperref}
\usepackage{booktabs}
\usepackage{tikz}
\usepackage{lipsum}
\usepackage{amssymb}
\usepackage{epsfig,mathrsfs,amsthm,graphicx,float,color}

\makeatletter
\DeclareFontFamily{OMX}{MnSymbolE}{}
\DeclareSymbolFont{MnLargeSymbols}{OMX}{MnSymbolE}{m}{n}
\SetSymbolFont{MnLargeSymbols}{bold}{OMX}{MnSymbolE}{b}{n}
\DeclareFontShape{OMX}{MnSymbolE}{m}{n}{
    <-6>  MnSymbolE5
   <6-7>  MnSymbolE6
   <7-8>  MnSymbolE7
   <8-9>  MnSymbolE8
   <9-10> MnSymbolE9
  <10-12> MnSymbolE10
  <12->   MnSymbolE12
}{}
\DeclareFontShape{OMX}{MnSymbolE}{b}{n}{
    <-6>  MnSymbolE-Bold5
   <6-7>  MnSymbolE-Bold6
   <7-8>  MnSymbolE-Bold7
   <8-9>  MnSymbolE-Bold8
   <9-10> MnSymbolE-Bold9
  <10-12> MnSymbolE-Bold10
  <12->   MnSymbolE-Bold12
}{}

\let\llangle\@undefined
\let\rrangle\@undefined
\DeclareMathDelimiter{\llangle}{\mathopen}%
                     {MnLargeSymbols}{'164}{MnLargeSymbols}{'164}
\DeclareMathDelimiter{\rrangle}{\mathclose}%
                     {MnLargeSymbols}{'171}{MnLargeSymbols}{'171}
\makeatother

\usepackage{mathtools}

\usepackage{physics}
\usepackage{subfigure}
\usepackage[ruled,vlined]{algorithm2e}
\def\<{\langle}\def\>{\rangle}
\def\bbra#1{\llangle#1\rvert}
\def\kett#1{\lvert#1\rrangle}
\def\Choi{\mathsf{Choi}}

\theoremstyle{plain}

\theoremstyle{definition}

\theoremstyle{remark}

\begin{document}

\title{Efficient tensor networks for control-enhanced quantum metrology}

\author{Qiushi Liu}
\email{qsliu@cs.hku.hk}
\affiliation{QICI Quantum Information and Computation Initiative, Department of Computer Science, School of Computing and Data Science, The University of Hong Kong, Pokfulam Road, Hong Kong, China}
\orcid{0000-0001-7352-8840}
\author{Yuxiang Yang}
\email{yuxiang@cs.hku.hk}
\affiliation{QICI Quantum Information and Computation Initiative, Department of Computer Science, School of Computing and Data Science, The University of Hong Kong, Pokfulam Road, Hong Kong, China}
\orcid{0000-0002-0531-8929}
\maketitle

\begin{abstract}
Optimized quantum control can enhance the performance and noise resilience of quantum metrology. However, the optimization quickly becomes intractable when multiple control operations are applied sequentially.  In this work, we propose efficient tensor network algorithms for optimizing strategies of quantum metrology enhanced by a long sequence of control operations. Our approach covers a general and practical scenario where the experimenter applies $N-1$ interleaved control operations between $N$ queries of the channel to estimate and uses no or bounded ancilla. Tailored to different experimental capabilities, these control operations can be generic quantum channels or variational unitary gates. Numerical experiments show that our algorithm has a good performance in optimizing the metrological strategy for as many as $N=100$ queries. In particular, our algorithm identifies a strategy that can outperform the state-of-the-art strategy when $N$ is finite but large.
\end{abstract}

\section{Introduction}
Quantum metrology \cite{giovannetti2004quantum,Giovannetti2006PRL,Degen17RMP} can boost the precision of parameter estimation by exploiting quantum resources. Identification of optimal strategies for large-scale quantum metrology, however, is a daunting task suffering from the curse of dimensionality. Such challenges, for example, arise in estimating physical parameters encoded in many-body quantum systems \cite{Huang23PRL,Chu23PRLstrong,Chabuda2020NC,Yin2024heisenberglimited,Dutkiewicz2024advantageofquantum} or many queries to an unknown quantum channel \cite{Fujiwara2008,Escher2011,Demkowicz-Dobrzanski2012,Kolodynski_2013NJP,Demkowicz14PRL,Zhou2021PRXQ,Kurdzialek23PRL}. In this work we focus on the quantum channel estimation problem.

It is a long-standing problem that, given $N$ queries to a quantum channel $\mathcal E_\theta$ carrying the parameter of interest $\theta$, what is the ultimate precision limit for estimating $\theta$ and how one could identify the optimal strategy that achieves the highest precision. For single-parameter estimation, precision bounds attainable by quantum error correction (QEC) \cite{Demkowicz-Dobrza2017PRX,Zhou2018NC,Zhou2021PRXQ} are known in the asymptotic limit ($N \rightarrow \infty$) and numerical algorithms for optimal strategies \cite{Altherr21PRL,Liu23PRL} have been designed for small $N$ (e.g., $N \le 5$). These results, however, are under the assumption that unbounded ancilla is available. Meanwhile, little is known about the metrological performance for a large but finite $N$. On the other hand, in real-world experiments, quantum channel estimation for larger $N$ with no or limited ancilla (for example, $N=8$ in Ref.~\cite{Hou2019PRL}) can be demonstrated with identical-control-enhanced sequential use of quantum channels \cite{Hou2019PRL,Hou2021PRL}. Obviously, a gap exists between the theoretical results and the experimental requirement. The reason is that we lack an efficient way to tackle such problems for large $N$.

A fruitful approach to the classical optimization for large-scale quantum problems, mainly in many-body physics, is the tensor network formalism \cite{ORUSAnnals,Bridgeman_2017}. Tensor networks provide an efficient way to classically store, manipulate and extract certain relevant information of quantum systems with limited many-body entanglement. In particular, this formalism has been used in quantum metrology for many-body systems to optimize the estimation precision and the corresponding input probe state \cite{Chabuda2020NC}.

It is natural to ask whether tensor networks can be used for optimizing a sequential strategy, where $N$ channels can be applied one after another, with interleaved control operations. This is possible by our intuition, as the power of tensor networks for addressing the spatial complexity is similarly applicable for handling the temporal complexity. For example, tensor networks have been used for quantum process tomography \cite{Torlai2023NC} and machine learning of non-Markovian processes \cite{Guo20PRA}.

\begin{figure} [!htbp]
\centering
\subfigure[Arbitrary control.]{\includegraphics[height=0.095\textwidth]{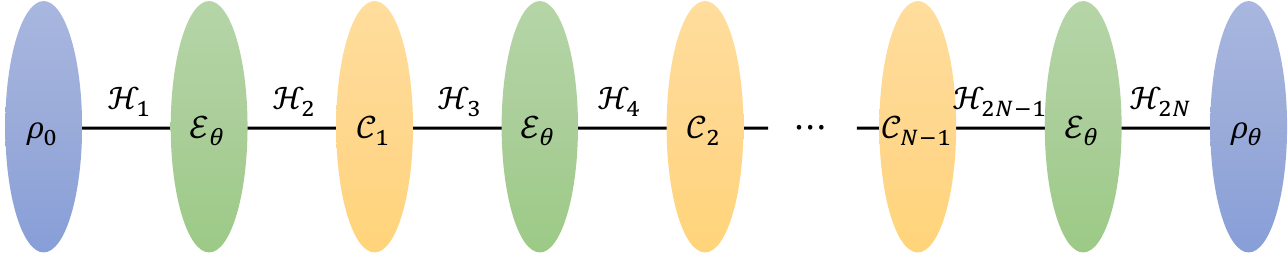}\label{fig:general setup}} \\
\subfigure[Identical control.]{\includegraphics[height=0.197\textwidth]{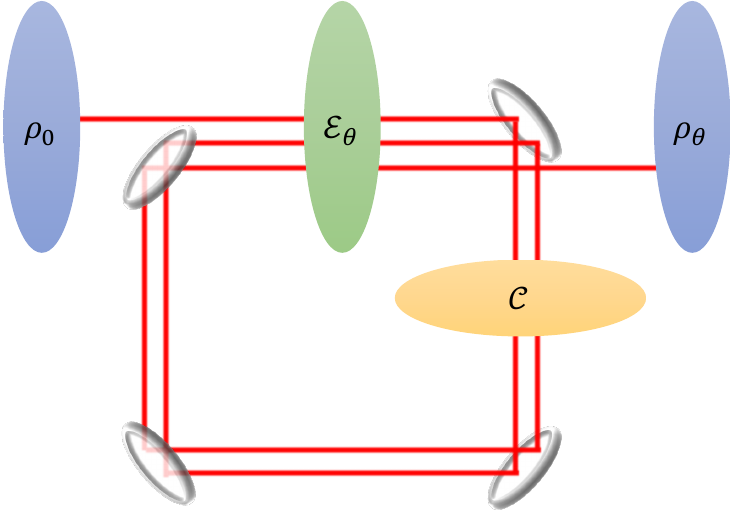}\label{fig:identical control setup}}
\caption{Control-enhanced sequential strategy for estimating $\theta$ from $N$ queries to $\mathcal E_\theta$, by interleaving $N-1$ control operations. Each control operation for $i=1,\dots,N-1$ is allowed to be an arbitrary quantum channel (CPTP map) $\mathcal C_i$. We can alternatively require all the control operations to be identical $\mathcal C_i = \mathcal C$, facilitating simpler experimental demonstration as in (b), taking $N=3$ with two identical control operations for an example. Note that this formalism can be readily generalized to the case with bounded ancilla.}
\label{fig:setup}
\end{figure}

In this work, we develop an efficient tensor network approach to optimizing control-enhanced sequential strategies for quantum channel estimation with bounded memory (see Fig.~\ref{fig:setup}). Tailored to the experimental capability and programmability of quantum sensors, we design various algorithms based on tensor networks, for applying control operations that are either allowed to be arbitrary or restricted to identical quantum channels. Furthermore, by combining the tensor network with a variational circuit ansatz, we also cover the scenario where the control operations are restricted to unitary channels specified by possibly different or identical variational parameters. Formulating the problem fully within the matrix product operator (MPO) formalism, we apply the alternating optimization iteratively. The computational complexity for each round of the optimization is $O(N)$, where the input state and the $N-1$ control operations are updated in each round. 
Our approach can efficiently optimize the estimation of qubit channels for $N=100$, and outperform the state-of-the-art asymptotically optimal QEC approach \cite{Demkowicz-Dobrza2017PRX,Zhou2018NC,Zhou2021PRXQ} for finite $N$.

\section{Framework} \label{sec:framework}
We start with some notations. We denote by $\mathcal L(\mathcal H)$ the set of linear operators on a finite-dimensional Hilbert space $\mathcal H$, and $\mathcal L[\mathcal L(\mathcal H_1), \mathcal L(\mathcal H_2)]$ is the set of linear maps from $\mathcal L(\mathcal H_1)$ to $\mathcal L(\mathcal H_2)$. For a positive semidefinite operator $A$, we simply write $A\ge 0$.

\subsection{Optimizing quantum Fisher information}

A fundamental task in quantum metrology is to attain the ultimate precision limit of estimating some parameter $\theta$, given many copies of a parametrized quantum state $\rho_\theta \in \mathcal L(\mathcal H)$. It is well known that the mean squared error (MSE) $\delta \theta^2$ for an unbiased estimator is bounded, according to the quantum Cramér-Rao bound (QCRB) \cite{helstrom1976quantum,holevo2011probabilistic,Braunstein1994PRL}
\begin{equation}
    \delta \theta^2 \ge \frac{1}{\nu F^Q(\rho_\theta)},
\end{equation}
where $F^Q(\rho_\theta)$ is the quantum Fisher information (QFI) and $\nu$ is the number of measurements that can be repeated. Importantly, the quantum Cramér-Rao bound is achievable for single-parameter estimation in the limit of $\nu \rightarrow \infty$. Therefore, the QFI is a key figure of merit in quantum metrology, and it is desirable to identify an optimal strategy that is most sensitive to the parameter acquisition and thus yields an output state $\rho_\theta$ that maximizes the QFI.

The problem formulation in this work can be illustrated in Fig.~\ref{fig:setup}. Given $N$ queries to a parametrized quantum channel $\mathcal E_\theta$, we would like to maximize the QFI of the output state $\rho_\theta$, by preparing an input state $\rho_0$ and inserting a sequence of control operations $\{\mathcal C_i\}_{i=1}^{N-1}$, where each $\mathcal C_i$ is a completely positive trace-preserving (CPTP) map. We can also restrict the control operations to be unitary channels $U(\boldsymbol{\phi}_i)$, choosing a variational circuit ansatz for simpler experimental demonstration. Each $\mathcal E_\theta \in \mathcal L[\mathcal L(\mathcal H_{2i-1}),\mathcal L(\mathcal H_{2i})]$ , for $i=1,\dots,N$, and each $\mathcal C_j \in \mathcal L[\mathcal L(\mathcal H_{2j}),\mathcal L(\mathcal H_{2j+1})]$, for $j=1,\dots,N-1$. For simplicity, we assume that the dimension of any $\mathcal H_i$ is the same $d=\mathsf{dim}(\mathcal H_i)$. We then have the output state
\begin{equation} \label{eq:output state}
    \rho_\theta = \Lambda_\theta(\rho_0) = \mathcal E_\theta \circ \mathcal C_{N-1} \circ \cdots \circ \mathcal C_2 \circ \mathcal E_\theta \circ \mathcal C_1 \circ \mathcal E_\theta (\rho_0).
\end{equation}
An optimal strategy is therefore a choice of $\rho_0$ and $\{\mathcal C_i\}$ such that the QFI $F^Q(\rho_\theta)$ is maximal. We assume that no ancilla is used in Fig.~\ref{fig:setup}, but this formulation also applies to the case where an ancilla space of fixed dimension is available, as one can replace $\mathcal E_\theta$ by $\mathcal E_\theta \otimes \mathcal I_A$ in the formulation (where $A$ denotes the ancilla space).

By definition, the QFI of a state $\rho_\theta$ is given by $F^Q(\rho_\theta)=\Tr(\rho_\theta L_{\theta}^2)$, where the symmetric logarithmic derivative (SLD) $L_{\theta}$ is the solution to $2\dot \rho_\theta = \rho_\theta L_{\theta} + L_{\theta} \rho_\theta$, having denoted the derivative of $X$ with respect to $\theta$ by $\dot{X}$. Computing the QFI by this definition, however, is often challenging when the optimization of metrological strategies is concerned. Instead, we can express the QFI $F^Q(\rho_\theta)$ as such a maximization problem \cite{macieszczak2013quantum,Macieszczak_2014NJP}:
\begin{equation}
    F^Q(\rho_\theta) = \sup_X \left[2\Tr(\dot{\rho_\theta} X) - \Tr(\rho_\theta X^2)\right],
\end{equation}
where $X$ is a Hermitian operator. One can check that the optimal solution for $X$ is exactly the SLD $L_\theta$. By combining Eq.~(\ref{eq:output state}), the maximal QFI obtained from $N$ uses of $\mathcal E_\theta$ is given by a multivariate maximization:
\begin{multline}\label{eq:QFI multivariate maximization}
     F^{(N)}(\mathcal E_\theta) = \sup_{\rho_0,\{\mathcal C_i\}} F^Q(\rho_\theta) \\= \sup_{\rho_0,\{\mathcal C_i\},X} \left\{2\Tr\left[ \dot{\Lambda}_\theta(\rho_0) X \right] - \Tr\left[\Lambda_\theta(\rho_0) X^2\right]\right\}.
\end{multline} 

Finding a global maximum for Eq.~(\ref{eq:QFI multivariate maximization}) can be a daunting task. However, considering that $-\left\{2\Tr\left[ \dot{\Lambda}_\theta(\rho_0) X \right] - \Tr\left[\Lambda_\theta(\rho_0) X^2\right]\right\}$ is a convex function of each variable in $\{\rho_0, \{\mathcal C_i\}, X\}$ while other variables are fixed, in practice we could use an \emph{alternating optimization} method: we search for an optimal solution for each variable by convex optimization, while keeping the other variables fixed, and iteratively repeat the procedure for optimizing all variables in multiple rounds. We remark that similar ideas have also been used for optimizing metrological strategies in many-body systems \cite{Chabuda2020NC} or with noisy measurements \cite{Len2022NC}. To implement the optimization in an efficient way for fairly large $N$, we will introduce the MPO formalism in Section \ref{sec:MPO}.

\subsection{Choi operator formalism} \label{sec:Choi formalism}
To facilitate the mapping of the optimization problem Eq.~(\ref{eq:QFI multivariate maximization}) into a tensor network, we first work on the quantum comb formalism \cite{gutoski2007toward,Chiribella2008PRL,Chiribella2009PRA} and characterize the quantum states and multi-step quantum processes by Choi operators \cite{choi1975completely} in a unified fashion. The Choi operator of a quantum state $\rho$ is still $\rho$ itself, and the Choi operator of a quantum channel $\mathcal E \in \mathcal L[\mathcal L(\mathcal H_{\mathrm{in}}), \mathcal L(\mathcal H_{\mathrm{out}})]$ is a positive semidefinite operator in $\mathcal L(\mathcal H_{\mathrm{out}} \otimes \mathcal H_{\mathrm{in}})$
\begin{equation}
    E = \Choi(\mathcal E) := \mathcal E \otimes \mathcal I(\kett I \bbra I), 
\end{equation}
where $\kett I=\sum_i \ket{i} \ket{i}$ is an unnormalized maximally entangled state and $\{\ket{i}\}_i$ is the computational basis of $\mathcal H_{\mathrm{in}}$. One can easily check that $\Tr_{\mathrm{out}} E= I_{\mathrm{in}}$. $\mathcal H_{\mathrm{in}}$ and $\mathcal H_{\mathrm{out}}$ can take multiple input-output pairs of a certain causal order, corresponding to a multi-step quantum process. The composition of quantum processes $\mathcal A$ and $\mathcal B$ is then characterized by the link product \cite{Chiribella2008PRL,Chiribella2009PRA} of Choi operators $A\in\mathcal L\left(\bigotimes_{a\in \mathsf A}\mathcal H_a\right)$ and $B\in\mathcal L\left(\bigotimes_{b\in \mathsf B}\mathcal H_b\right)$, defined by
\begin{equation}
    A*B := \Tr_{\mathsf A\cap\mathsf B}\left[\left(I_{\mathsf B\backslash\mathsf A}\otimes A^{T_{\mathsf A\cap\mathsf B}}\right)\left(B\otimes I_{\mathsf A\backslash\mathsf B}\right)\right],
\end{equation}
where $T_i$ denotes the partial transpose on $\mathcal H_i$, $\mathsf A\cap\mathsf B :=\{i\mid i \in \mathsf A,i \in \mathsf B\}$, $\mathsf A\backslash\mathsf B:=\{i\mid i \in \mathsf A,i \notin \mathsf B\}$, and $\mathcal H_{\mathsf X}$ denotes $\bigotimes_{i\in \mathsf X} \mathcal H_{i}$.

With the Choi-Jamiołkowski isomorphism, we denote the Choi operators of $\mathcal E_\theta \in \mathcal L[\mathcal L(\mathcal H_{2i-1}),\mathcal L(\mathcal H_{2i})]$ and $\mathcal C_i \in \mathcal L[\mathcal L(\mathcal H_{2i}),\mathcal L(\mathcal H_{2i+1})]$ by $E_\theta:=\Choi(\mathcal E_\theta)$ and $C_i:=\Choi(\mathcal C_i)$. Now we can reformulate Eq.~(\ref{eq:QFI multivariate maximization}) as
\begin{multline} \label{eq:QFI multivariate maximization Choi}
    F^{(N)}(\mathcal E_\theta) \\= \sup_{\rho_0,\{C_i\},X} \left\{2\Tr\left[\left(\frac{dE_\theta^{\otimes N}}{d\theta}*\bigotimes_{i=1}^{N-1} C_i \otimes \rho_0\right)X\right]\right.\\
    \left.- \Tr\left[\left(E_\theta^{\otimes N}*\bigotimes_{i=1}^{N-1} C_i \otimes \rho_0\right)X^2\right]\right\},
\end{multline}
where $\rho_0 \ge 0$, $\Tr(\rho_0)=1$, $C_i \ge 0$, and $\Tr_{2i+1} C_i = I_{2i}$ for any $i=1,\dots,N-1$. For ease of notation we define
\begin{equation} \label{eq:f1}
    f_1 := \Tr\left[\left(\frac{dE_\theta^{\otimes N}}{d\theta}*\bigotimes_{i=1}^{N-1} C_i \otimes \rho_0\right)X\right]
\end{equation}
and
\begin{equation} \label{eq:f2}
    f_2 := \Tr\left[\left(E_\theta^{\otimes N}*\bigotimes_{i=1}^{N-1} C_i \otimes \rho_0\right)X^2\right].
\end{equation}
A main advantage of working with Choi operators is that, it facilitates the optimization of each variable in $\{\rho_0,\{C_i\},X\}$. While other variables are fixed in the alternating optimization, the optimal solution for each variable in a convex set can be identified either by simple diagonalization or by semidefinite programming (SDP), as we will see later. A naive implementation of the optimization, however, concerns a Hilbert space whose dimension exponentially grows with $N$. It is essential to compute and optimize $2f_1-f_2$ efficiently in a tensor network formalism, where all matrices we need to deal with are in the form of MPO.

\subsection{Matrix product operators for efficient optimization} \label{sec:MPO}
MPO is an efficient way to store and manipulate a high-dimensional matrix/tensor. As a simple example, if we would like to store the information of $E_\theta^{\otimes N}$ (where each $E_\theta$ is a $d^2 \times d^2$ matrix), we do not need to explicitly store a $d^{2N} \times d^{2N}$ matrix. Instead, it is only necessary to store the information of  $E_\theta$, so that we can retrieve each component of $E_\theta^{\otimes N}$ efficiently. To be more specific, we write 
\begin{multline}
    E_\theta = \\ \sum_{\alpha_{2i}\alpha_{2i-1}\beta_{2i}\beta_{2i-1}} (E_\theta)^{\alpha_{2i}\alpha_{2i-1}}_{\beta_{2i}\beta_{2i-1}}\ketbra{\alpha_{2i}\alpha_{2i-1}}{\beta_{2i}\beta_{2i-1}},
\end{multline}
which is an operator in $\mathcal L(\mathcal H_{2i} \otimes \mathcal H_{2i-1})$. We can thus retrieve any component $(E_\theta^{\otimes N})^{\alpha_{2N}\dots\alpha_2\alpha_1}_{\beta_{2N}\dots\beta_2\beta_1}$ efficiently with $O(N)$ complexity, by calculating a product $(E_\theta)^{\alpha_{2N}\alpha_{2N-1}}_{\beta_{2N}\beta_{2N-1}}\cdots (E_\theta)^{\alpha_{2}\alpha_{1}}_{\beta_{2}\beta_{1}}$. 

\begin{figure} [!htbp]
\centering
\subfigure[Tensor network for $f_2$.]{\includegraphics[width=0.48\textwidth]{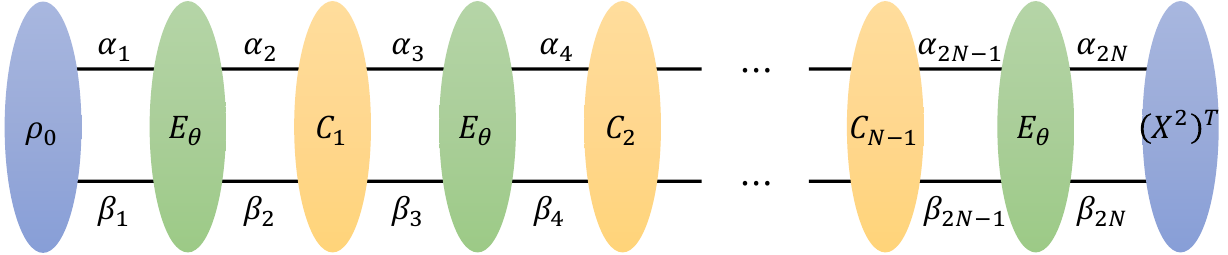}\label{fig:f_2}} \\
\subfigure[Tensor network for $f_1$.]{\includegraphics[width=0.48\textwidth]{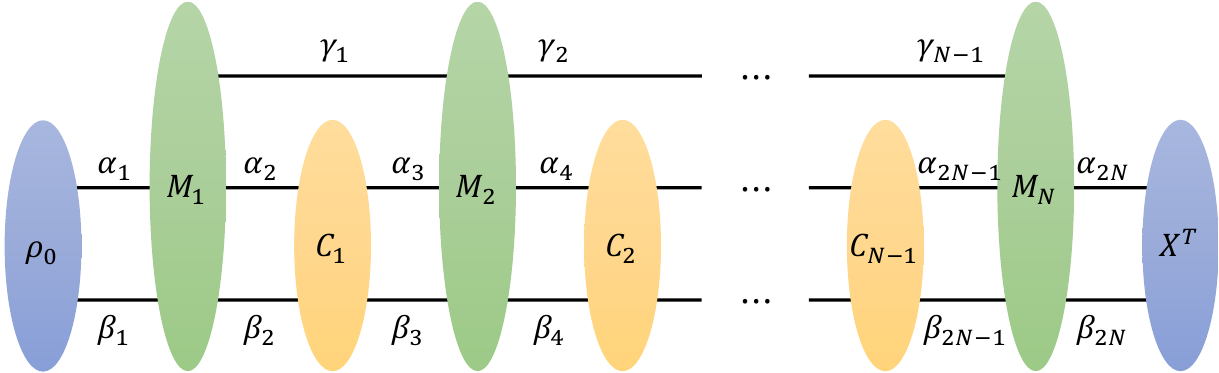}\label{fig:f_1}}
\caption{Tensor networks for computing $f_2$ in Eq.~(\ref{eq:f2}) and $f_1$ in Eq.~(\ref{eq:f1}) for evaluating the QFI. In (b), $M_1,\dots,M_N$ are matrices containing the information about both $E_\theta$ and $\dot E_\theta$.}
\label{fig:tensor networks for f_1 and f_2}
\end{figure}

The power of MPO is definitely not limited to this simple case of expressing the tensor product of matrices. In fact, MPO ensures the efficiency when the dimension of shared tensor indices (called the bond dimension) can be kept relatively low during tensor contraction. We will see that we can evaluate and optimize $f_1$ and $f_2$ (and thus the QFI) in the MPO formalism with a low bond dimension. First, $f_2$ defined by Eq.~(\ref{eq:f2}) is expressed as
\begin{multline}
    f_2 = (X^2)^{\beta_{2N}}_{\alpha_{2N}} (E_\theta)^{\alpha_{2N} \alpha_{2N-1}}_{\beta_{2N} \beta_{2N-1}}\cdots \\(C_1)^{\alpha_3 \alpha_2}_{\beta_3 \beta_2}(E_\theta)^{\alpha_2 \alpha_1}_{\beta_2 \beta_1}(\rho_0)^{\alpha_1}_{\beta_1},
\end{multline}
having taken the convention of Einstein summation, where repeated indices are implicitly summed over. This can be schematically illustrated in Fig.~\ref{fig:f_2}. Each leg represents an index, and a connected line represents a summation over the corresponding index. Computing $f_2$ requires $O(N)$ complexity.

Calculating $f_1$ given by Eq.~(\ref{eq:f1}) is a bit more involved, as the derivative $dE_\theta^{\otimes N}/d\theta$ needs to be evaluated. One may simply use $dE_\theta^{\otimes N}/d\theta=\sum_{i=1}^N E_\theta^{\otimes i-1} \otimes \dot E_\theta \otimes E_\theta^{\otimes N-i}$, but this introduces an $O(N^2)$ computational overhead, which is unfavorable when $N$ is large. To circumvent this problem, we draw a technique for expressing a sum of local terms as an MPO \cite{Crosswhite08PRA,mcculloch2008infinite,Hubig17PRB} from many-body Hamiltonian representation. In Appendix \ref{app:MPO derivative} we show that $dE_\theta^{\otimes N}/d\theta$ can be written as an MPO with a bond dimension (corresponding to the $\gamma$ label) of $2$:
\begin{multline}\label{eq:MPO derivative}
    \frac{dE_\theta^{\otimes N}}{d\theta} = (M_N)^{\alpha_{2N}\alpha_{2N-1}}_{\beta_{2N}\beta_{2N-1}\gamma_{N-1}}\cdots \\(M_2)^{\alpha_{4}\alpha_{3}}_{\beta_{4}\beta_{3}\gamma_2\gamma_1} (M_1)^{\alpha_{2}\alpha_{1}}_{\beta_{2}\beta_{1}\gamma_{1}},
\end{multline}
where (for simplicity, we only explicitly focus on the $\gamma$ indices for the two-dimensional bonds connecting $M_1,\dots,M_N$)
\begin{multline}
    (M_N)_1 = (M_1)_2 = E_\theta,\ (M_N)_2 = (M_1)_1 = \dot E_\theta,\\
    (M_i)_{11} = E_\theta,\ (M_i)_{12}=\dot E_\theta,\\ (M_i)_{21}=0,\ (M_i)_{22}=E_\theta,\\
    \forall i=2,\dots,N-1.
\end{multline}
With the MPO representation of $E_\theta^{\otimes N}/d\theta$, we can easily see that $f_1$ can be computed as the contraction of a tensor network with $O(N)$ complexity at a small cost of bond dimension of $2$, schematically illustrated in Fig.~\ref{fig:f_1}. We defer the contraction of $f_1$ and $f_2$ in the bounded ancilla case to Appendix \ref{app: f1 and f2 with ancilla}.

The efficiency of a tensor network largely depends on the contraction order of indices. The key idea is to leave open legs as few as possible during the contraction. For our purpose this is simple: we can follow a quantum circuit time flow order (or the reverse order), and the number of open legs would not accumulate. A more detailed complexity analysis of tensor contractions during the optimization will be given in Section \ref{sec:optimization procedure}.

\subsection{Optimization procedure: arbitrary control} \label{sec:optimization procedure}
With the tensor networks for $f_1$ and $f_2$, we can now efficiently optimize the QFI $F^{(N)}(\mathcal E_\theta) = \sup_{\rho_0,\{C_i\},X} (2f_1-f_2)$ by alternating optimization.

We start with an initialization of the input state $\rho_0$ and the control operations $\{C_i\}$. The initialization can be random, or we can start from a previously known strategy which has a relatively good performance. We can use the tensor network for $f_2$ without the last tensor $(X^2)^T$ [denoted by $f_2\backslash (X^2)^T$] to compute the output state $\rho_\theta$, and use the tensor network for $f_1$ without the last tensor $X^T$ (denoted by $f_1\backslash X^T$) to compute the derivative $\dot \rho_\theta$. We then calculate the SLD of the output state \cite{Braunstein1994PRL}
\begin{equation}
    L_\theta = \sum_{jk}\frac{2\mel{j}{\dot \rho_\theta}{k}}{\mel{j}{\rho_\theta}{j}+\mel{k}{\rho_\theta}{k}}\ketbra{j}{k},
\end{equation}
where $\rho_\theta = \sum_j \mel{j}{\rho_\theta}{j}\ketbra{j}{j}$ for an orthonormal eigenbasis $\{\ket{j}\}_j$ of $\rho_\theta$. Thus $L_\theta$ is the solution for the optimal value of $X$ so far.

Now to update the input state $\rho_0$, we contract the tensor network $f_1$ without the first tensor $\rho_0$ (denoted by $f_1\backslash\rho_0$). Similarly, we compute $f_2\backslash\rho_0$. Our objective is thus to maximize
\begin{multline}
    (\rho_0)^{\alpha_1}_{\beta_1}(2f_1\backslash \rho_0-f_2\backslash \rho_0)^{\alpha_1}_{\beta_1}\\
    = \Tr\left[\rho_0(2f_1\backslash \rho_0-f_2\backslash \rho_0)^T\right].
\end{multline}
The updated $\rho_0'$ can be identified as $\rho_0'=\dyad{\psi_0'}$, where $\ket{\psi_0'}$ is a normalized eigenvector of $(2f_1\backslash \rho_0-f_2\backslash \rho_0)^T$ associated with the maximal eigenvalue. Due to the convexity of QFI, the optimal input state can always be chosen as a pure state.

For each control operation $C_i$, we contract $f_1\backslash C_i$ and $f_2 \backslash C_i$ (with all the tensors updated so far), and compute $2 f_1\backslash C_i-f_2\backslash C_i$. Without loss of generality, let us take $C_1$ for example. The optimization of $C_1$ can be formulated as an SDP:
\begin{multline}\label{eq:control optimization sdp} 
        \max_{C_1}\left[ (C_1)^{\alpha_3 \alpha_2}_{\beta_3 \beta_2}(2 f_1\backslash C_1-f_2\backslash C_1)^{\alpha_3 \alpha_2}_{\beta_3 \beta_2}\right],\\
        \mathrm{s.t.}\ C_1 \ge 0, \Tr_3 C_1 = I_2.
\end{multline}
We can apply similar optimization for $C_2,\dots,C_{N-1}$. It is important that each $C_i$ is updated by locally solving an optimization problem to circumvent the exponential overhead of a global optimization. Such an optimization procedure is motivated by the well-known algorithm for solving ground states of many-body Hamiltonians via the density matrix renormalization group (DMRG) \cite{WhiteDMRG92PRL,WhiteDMRG93PRB}, wherein a matrix product state instead of an MPO representation of control operations is locally and iteratively updated.

It is noteworthy that the constraints in Eq.~(\ref{eq:control optimization sdp}) can be adapted when the control operations are restricted. For instance, if we only allow for entanglement breaking channels \cite{Horodecki03EntanglementBreaking} as control, then each Choi operator $C_i$ should be optimized over the convex set of (unnormalized) separable states. For the ancilla-free control over a qubit system, the separability of $C_i$ can be fully characterized by the positive partial transpose (PPT) criterion \cite{Peres96PRL,HORODECKI19961}, i.e., we require $C_i^{T_{2i}} \ge 0$, where ${T_{2i}}$ denotes the partial transpose on $\mathcal H_{2i}$. In general for higher-dimensional quantum control operations, the PPT criterion is a necessary but not always sufficient condition for the separability of the Choi operator. In fact, characterizing separability is an NP-hard problem \cite{Gurvits03Classical}. Nevertheless, the optimization with separability constraints can be approximately solved by a sequence of SDPs with convergence guarantee \cite{Doherty02PRL,Doherty04PRA,Yu2021NC,Yu2022PRXQuantum}.

We can repeat the procedure above iteratively in an outer loop of optimizing $\{X,\rho_0,\{C_i\}\}$, until the required convergence is satisfied. Finally, we have both the optimized QFI $2f_1-f_2$ and a concrete strategy to attain it, including a probe state and a set of control operations.

We conclude this section with a complexity analysis of the tensor contraction in each round of the optimization. Assume that $\mathcal E_\theta$ acts on a $d$-dimensional system. To update $N+1$ variable tensors $\left\{X,\rho_0,\{C_i\}_{i=1}^{N-1}\right\}$, we need to contract the tensor network $O(N)$ times. Each contraction of the tensor network for updating a variable requires $O(Nd^4)$ operations by sequentially following a quantum circuit time order or the reverse order. Here the factor $d^4$ arises from the $d^2$ values that open indices run over and the $d^2$ values that connected indices can take, while contracting two tensors each step. In particular, it is important to note that the intermediate results of each contraction can be saved for later use. For example, while updating $C_1$, one should keep track of all the intermediate contractions with respect to $C_2,\dots,C_{N-1}$ (in the reverse order). Thus in the next step of updating $C_2$, the contraction with respect to $C_3,\dots,C_{N-1}$ can be reused and one only needs to contract the tensor network with the updated $C_1$ and save the result, which admits $O(d^4)$ computational complexity. Similar arguments hold for updating the remaining tensors. Therefore, each round of the optimization requires $O(Nd^4)$ operations. 

\subsection{Using identical control operations}
In many practical scenarios, for example, for quantum sensors with limited programmability, it is favorable to apply the same control operation at each step. To this end, we propose an alternative approach to optimizing the control operations which are required to be identical, inspired by the infinite MPO (iMPO) \cite{mcculloch2008infinite,Cincio13PRL,Corboz16PRB}. Similar techniques have been used in Ref.~\cite{Chabuda2020NC} for quantum metrology of many-body systems in the infinite particle limit. Note that in this work we employ this technique for finding identical control operations for a finite $N$.

When the control operations are constrained to be identical, the tensor network optimization becomes a highly nonlinear problem, and we employ a trick to globally update all the control based on the solution from local optimization. First, we need to initialize all the identical control operations $C_i=C$. With a little abuse of the notation, we denote by $2f_1(C)-f_2(C)$ the contraction result of $2f_1-f_2$ with identical control operations $C_i=C$. In each iteration step,  we pick a control operation $C_i$ at random (for example, $C_1$) and still need to solve an SDP as in Eq.~(\ref{eq:control optimization sdp}), keeping all the tensors fixed except for $C_1$. The optimal solution is denoted by $\tilde C_i$. Instead of directly choosing $\tilde C_i$ to be the updated control, we take a convex combination \cite{Corboz16PRB}
\begin{equation}
    C_i'(\lambda) = \sin^2(\lambda \pi) \tilde C_i + \cos^2(\lambda \pi) C_i,
\end{equation}
for some $\lambda \in [0,0.5]$ and optimize the choice of $\lambda$ to maximize $2f_1(C)-f_2(C)$. In practice, it usually works well by performing a local optimization of the variable $\lambda$ (for example, using the bounded method \cite{Forsythe1977computer,Brent2013algorithms} in SciPy \cite{2020SciPy-NMeth}), since we are interested in the convergence result after a number of iterations.

We remark that $2f_1(C)-f_2(C)$ admits fast evaluation even for large $N$, due to the translation invariant part of the tensor network. Specifically, for evaluating $f_2(C)$, we reshape the tensors as matrices $(\mathbf{E}_\theta)_{\alpha_4\beta_4,\alpha_3\beta_3} = (E_\theta)^{\alpha_{4} \alpha_{3}}_{\beta_{4} \beta_{3}}$ and $(\mathbf{C})_{\alpha_3\beta_3,\alpha_2\beta_2} = (C)^{\alpha_{3} \alpha_{2}}_{\beta_{3} \beta_{2}}$. Then instead of contracting a sequence of tensors, we simply compute the power of a matrix
\begin{multline}
    \left[(\mathbf{E}_\theta \mathbf C)^{N-1}\right]_{\alpha_{2N}\beta_{2N},\alpha_{2}\beta_{2}} \\ = (E_\theta)^{\alpha_{2N} \alpha_{2N-1}}_{\beta_{2N} \beta_{2N-1}} C^{\alpha_{2N-1} \alpha_{2N-2}}_{\beta_{2N-1} \beta_{2N-2}} \cdots (E_\theta)^{\alpha_4 \alpha_3}_{\beta_4 \beta_3} C^{\alpha_3 \alpha_2}_{\beta_3 \beta_2},
\end{multline}
from which we can easily obtain $f_2(C)$. Analogously, to compute $f_1(C)$, we reshape $(M_i)^{\alpha_{2i}\alpha_{2i-1}}_{\beta_{2i}\beta_{2i-1}\gamma_{i}\gamma_{i-1}}$ as matrix $\mathbf{M}_{\alpha_{2i}\beta_{2i}\gamma_{i},\alpha_{2i-1}\beta_{2i-1}\gamma_{i-1}}$ for $i=2,\dots,N-1$, and compute the matrix power
\begin{multline}
    \left\{\left[\mathbf{M} (\mathbf C\otimes I)\right]^{N-2}\right\}_{\alpha_{2N-2}\beta_{2N-2}\gamma_{N-1},\alpha_{2}\beta_{2}\gamma_{1}} \\ = (M_{N-1})^{\alpha_{2N-2}\alpha_{2N-3}}_{\beta_{2N-2}\beta_{2N-3}\gamma_{N-1}\gamma_{N-2}} C^{\alpha_{2N-3} \alpha_{2N-4}}_{\beta_{2N-3} \beta_{2N-4}} \cdots \\ (M_{2})^{\alpha_{4}\alpha_{3}}_{\beta_{4}\beta_{3}\gamma_{2}\gamma_{1}} C^{\alpha_3 \alpha_2}_{\beta_3 \beta_2},
\end{multline}
from which $f_1(C)$ can be immediately obtained.

\subsection{Variational unitary control} 
Implementing an arbitrary CPTP control operation is sometimes challenging. To be more experiment-friendly, we can alternatively take a simple variational ansatz for the control (see Appendix \ref{app:variational ansatz}), and thus characterize each control by several variational parameters. In this case $C_i = \Choi\left[\mathcal U(\boldsymbol{\phi_i})\right]$ is the Choi operator of the variational circuit unitary $U(\boldsymbol{\phi}_i)$, where $\boldsymbol{\phi}_i$ is a vector of variational parameters for the $i$-th control operation, and $\mathcal U(\boldsymbol{\phi_i})(\rho)=U(\boldsymbol{\phi_i})\rho \left[U(\boldsymbol{\phi_i})\right]^\dagger$. 

Previously variational circuits have been mostly used for optimizing small-scale problems in quantum metrology \cite{Koczor_2020,Yang2020,Ma21IEEE,Meyer2021,Beckey22PRRvariational,Le2023variational,Altherr21PRL}. The tensor network formulation here, however, allows for optimizing a long sequence of variational control operations efficiently. Concatenating the variational ansatz with the tensor network algorithm, we employ a local update method if each control operation can be different, and a global update method if all the control operations are restricted to be identical.

\subsubsection{Local update: arbitrary variational parameters}
We first introduce the local update method, which is just a slight modification of the optimization procedure introduced in Section \ref{sec:optimization procedure}. Instead of solving an SDP for each control operation $C_i$, each time we fix $X$, $\rho_0$, and all the other control $\{U(\boldsymbol{\phi}_j)\}_{j \neq i}$, take the cost function $-\left[2f_1(\boldsymbol{\phi_i})-f_2(\boldsymbol{\phi_i})\right]$ as a function of $\boldsymbol{\phi}_i$, and tune the variational parameters $\boldsymbol{\phi}_i$ by gradient descent to minimize the cost function. The gradient estimation is both time-efficient and memory-efficient as each variational control is locally updated.

\subsubsection{Global update: identical variational parameters}
For a simpler experimental demonstration of quantum sensors, we can further restrict $\boldsymbol{\phi}_i=\boldsymbol{\phi}$ for all control operations, admitting a succinct circuit description. To update the variational parameters $\boldsymbol{\phi}$ in each iteration, we need to estimate the gradient with respect to $\boldsymbol{\phi}$:
\begin{multline}
    \nabla_{\boldsymbol{\phi}} (f_2-2f_1) \\
    = \nabla_{\boldsymbol{\phi}} C^{\alpha_{3} \alpha_{2}}_{\beta_{3} \beta_{2}}\sum_i (f_2\backslash C_i-2 f_1\backslash C_i)^{\alpha_{3} \alpha_{2}}_{\beta_{3} \beta_{2}},
\end{multline}
where $C_i=C=\Choi\left[\mathcal U(\boldsymbol{\phi})\right]$ for all $i=1,\dots,N-1$. Note that we can efficiently compute the summation over the control index $i$ by sharing intermediate contraction results in a similar fashion as in Section \ref{sec:optimization procedure}.

\section{Numerical experiments} \label{sec:applications}
In this section, we apply the algorithms introduced above to different noise models for $N=2$ to $N=100$ queries. For the first example, phase estimation with the bit flip noise can achieve the Heisenberg limit (HL) [$F^{(N)}(\mathcal E_\theta) \propto N^2$]. When $N$ is finite, our algorithm can outperform the state-of-the-art asymptotically optimal approach based on QEC \cite{Demkowicz-Dobrza2017PRX,Zhou2018NC,Zhou2021PRXQ} as well as other known strategies. For the second example of the amplitude damping noise, it is known a priori that the standard quantum limit (SQL) [$F^{(N)}(\mathcal E_\theta) \propto N$] can be attained with unbounded ancilla \cite{Zhou2021PRXQ}, but when the ancilla is bounded a theoretical analysis of the optimal metrological performance is lacking. Here, our algorithm can yield control strategies that significantly outperform a control-free one. Note that in this work we take the convention that signal comes after noise as in Ref.~\cite{Kurdzialek23PRL}, which is more interesting in some scenarios, where the HL can be achieved in noisy channel estimation, but the coefficient of the scaling is smaller than the noiseless case. As a final example of non-unitary encoding, we investigate the estimation of the direction angle of the dephasing noise by our algorithm.

To achieve faster convergence\footnote{In some cases it may also be helpful to add some perturbation (for example, an additional depolarizing noise decaying versus iterations) to the channel to estimate $\mathcal E_\theta$. We suspect that this may help circumvent the possible discontinuity of the QFI at the parameter value where the rank of the output state $\rho_\theta$ changes \cite{Dominik17PRA,Seveso_2020,zhou2019exact,Ye22PRA}.}, we may try starting from a known strategy (e.g., QEC), or initialize the configuration in an adaptive way: we choose the input state and control operations based on the optimized strategy for $N$ queries, as a starting point for optimizing the strategy with $N+1$ queries. The algorithm typically converges well in around $50$--$200$ iterations.

\subsection{Phase estimation with the bit flip noise}
We start with the extensively studied bit flip noise model, and compare our results with some existing works. In the ancilla-free scenario, our result matches a recent finding \cite{zhou2024limits} for achieving the SQL. Assisted by one ancilla, our result outperforms the asymptotically optimal QEC strategy \cite{Demkowicz-Dobrza2017PRX,Zhou2018NC,Zhou2021PRXQ} by using CPTP control, and outperforms the best known ancilla-free strategy by using unitary control. A summary of known results of the asymptotic QFI for comparison is given in Table \ref{tab:QFI scalings BF}.

\begin{table}[!htbp]
\caption{\label{tab:QFI scalings BF} Known scalings of the QFI for phase estimation with the bit flip noise by control-enhanced sequential strategies.}
\centering
\begin{tabular}{lll} \toprule
Control type & CPTP & Unitary \\
\midrule
No ancilla \cite{zhou2024limits} & $O(N^{1.5})$ & $\Theta(N)$ \\
Bounded ancilla & $\Theta(N^2)$ \cite{Kessler14PRL} & $\Theta(N)$ \cite{zhou2024limits}\\
Unbounded ancilla \cite{Kessler14PRL} & $\Theta(N^2)$ & $\Theta(N^2)$ \\
\bottomrule
\end{tabular}
\end{table}

We would like to estimate $\theta$ from $N$ queries to a quantum channel $\mathcal E_\theta$ described by the Kraus operators $\{K^\theta_i=U_Z(\theta) K^{(\mathrm{BF})}_i\}_{i=1}^2$, where $U_Z(\theta)=e^{-\mathrm i \theta \sigma_z/2}$ and
\begin{equation}
    K^{(\mathrm{BF})}_1 = \sqrt{1-p} I,\ K^{(\mathrm{BF})}_2 = \sqrt{p} \sigma_x,
\end{equation}
for $0\le p \le 1$. In numerical experiments, we take $p=0.1$ and the ground truth $\theta=\theta_0=1.0$.

\begin{figure}[!htbp]
    \centering
    \includegraphics[width=0.48\textwidth]
    {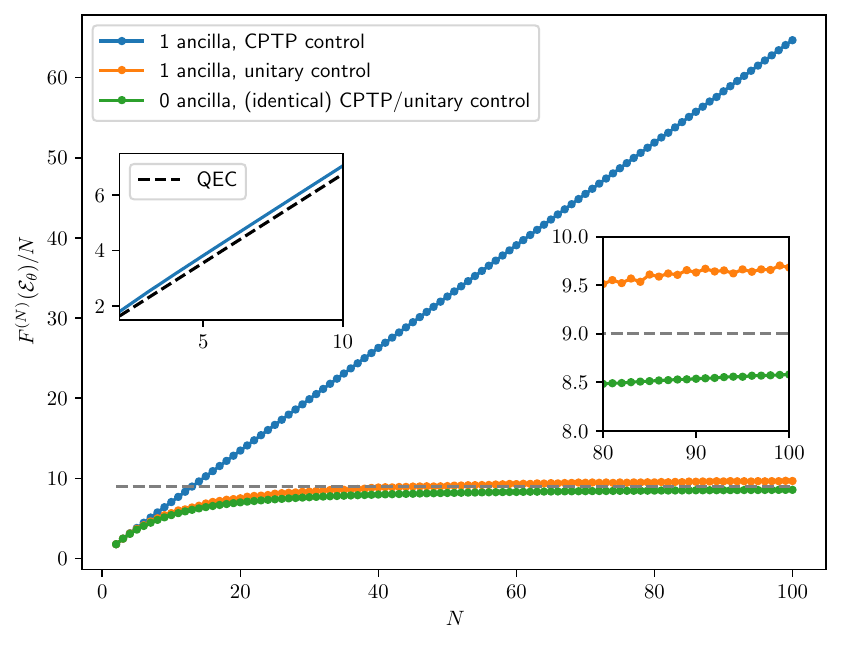}
    \caption{The normalized QFI $F^{(N)}(\mathcal E_\theta)/N$ versus the number of queries $N$ to $\mathcal E_\theta$ for phase estimation with the bit flip noise. The left inset figure shows that $F^{(N)}(\mathcal E_\theta)/N$ assisted by $1$ ancilla and CPTP control operations (solid blue line) can be higher than that obtained by an optimal QEC strategy (dashed black line) for $N=2$ to $10$. This advantage will become smaller when $N$ is larger (but still exists for $N=100$), which is expected as the QEC strategy is asymptotically optimal. The right inset shows that, with $1$ ancilla and unitary control (solid orange line), $F^{(N)}(\mathcal E_\theta)/N$ can surpass the the asymptotic limit of the ancilla-free strategy (dashed gray line) proposed in Ref.~\cite{zhou2024limits} when $N$ is large enough.}
    \label{fig:phase estimation BF comparison}
\end{figure}

We compare the normalized QFI $F^{(N)}(\mathcal E_\theta)/N$ obtained by different strategies versus $N$ for $N=2$ to $100$ queries to $\mathcal E_\theta$ in Fig.~\ref{fig:phase estimation BF comparison}. First, in the ancilla-free scenario, we apply the optimization algorithms using arbitrary/identical, CPTP/variational unitary control operations. In all these $4$ cases the obtained results coincide with each other: the growth of $F^{(N)}(\mathcal E_\theta)/N$ slows down as $N$ increases, which implies a standard quantum scaling. We further find out that the control operation $U_c$ output by our algorithm is very close to the inverse of the unitary encoding the signal at the ground truth $\theta=\theta_0$:
\begin{equation}
    U_c \approx U_Z(\theta_0)^\dagger.
\end{equation}
This result matches the theoretical finding in a recent work \cite[Theorem 5]{zhou2024limits}, which states that for this noise model the SQL can be achieved in the asymptotic limit by nearly reversing the signal encoding unitary, such that $\lim_{N\rightarrow\infty} F^{(N)}(\mathcal E_\theta)/N$ can be arbitrarily close to $\frac{1-p}{p}$ \cite{zhou2024limits}, depicted as a dashed gray line in Fig.~\ref{fig:phase estimation BF comparison} for comparison\footnote{It is worth noting that, however, exactly choosing $U_c = U_Z(\theta_0)^\dagger$ results in the discontinuity of the QFI, as the matrix rank of the output state $\rho_\theta$ changes at this point of $\theta=\theta_0$, and this can also cause the numerical instability of the algorithm.}. 

Next, we investigate the performance of the strategy assisted by one ancilla qubit. In this case, an asymptotically optimal QEC strategy requiring one ancilla is known \cite{Zhou2021PRXQ} and yields the HL $\lim_{N\rightarrow \infty} F^{(\mathrm{N})}(\mathcal E_\theta)/N^2 = (1-2p)^2$. However, the QEC is not optimal for a finite $N$ \cite{Kurdzialek23PRL,Altherr21PRL,Liu23PRL}. Fig.~\ref{fig:phase estimation BF comparison} shows that the QFI yielded by using arbitrary CPTP control can outperform that yielded by an optimal QEC strategy. To make a fair comparison, we apply $N-1$ QEC recovery operations between $\mathcal E_\theta$ and exclude the final recovery operation (which maps the state back to the codespace) at the end. We also make sure that QEC approximately reverses the effect of the signal unitary (up to a unitary $e^{\mathrm{-i}\epsilon \sigma_z /2}$ with a small $\epsilon=0.01$), otherwise the QFI obtained from QEC may significantly decrease due to the discontinuity. We illustrate the metrological performance gap between the QEC strategy and the tensor network algorithm for $N=2$ to $10$ in the left inset of Fig.~\ref{fig:phase estimation BF comparison}. When $N$ gets large (e.g., $N \approx 100$), the relative advantage becomes smaller but still exists. We expect that the relative gap would vanish in the asymptotic limit, as the QEC strategy is asymptotically optimal. Nevertheless, our algorithm provides a higher metrological sensitivity in real-world experiments with finite $N$. 

Moreover, by using variational unitary control operations with one ancilla qubit, our algorithm yields a normalized QFI $F^{(\mathrm{N})}(\mathcal E_\theta)/N$ higher than the theoretical limit $\frac{1-p}{p}$ of the best existing ancilla-free strategy \cite{zhou2024limits}, depicted by the right inset of Fig.~\ref{fig:phase estimation BF comparison} for $N=80$ to $100$. Our results indicate that, even if the QFI with unitary control and bounded ancilla can only achieve the SQL \cite{zhou2024limits} as in the ancilla-free case, the ancilla could still be useful by improving the coefficient of the scaling.

\subsection{Phase estimation with the amplitude damping noise}
With the amplitude damping noise, the channel to estimate $\mathcal{E}_\theta$ is characterized by the Kraus operators $\{K^\theta_i=U_Z(\theta) K^{(\mathrm{AD})}_i\}_{i=1}^2$, where
\begin{equation}
    K^{(\mathrm{AD})}_1 = \left(\begin{array}{cc}
        1 & 0\\
        0 & \sqrt{1-p}
  \end{array}\right),\ K^{(\mathrm{AD})}_2 = \left(\begin{array}{cc}
        0 & \sqrt{p}\\
        0 & 0
  \end{array}\right),
\end{equation}
for $0\le p \le 1$. In contrast to the previous example (i.e., the bit flip model), under amplitude damping noise, phase estimation cannot achieve the HL even with unbounded ancilla \cite{Fujiwara2008,Escher2011,Demkowicz-Dobrzanski2012}. A summary of known scalings of QFI is presented in Table \ref{tab:QFI scalings AD}.

\begin{table}[!htbp]
\caption{\label{tab:QFI scalings AD} Known scalings of the QFI for phase estimation with the amplitude damping noise by control-enhanced sequential strategies.}
\centering
\begin{tabular}{lll} \toprule
Control type & CPTP & Unitary \\
\midrule
No ancilla \cite{zhou2024limits} & $\Theta(1)$ & $O(1)$ \\
Bounded ancilla \cite{Demkowicz14PRL} & $O(N)$ & $O(N)$ \\
Unbounded ancilla \cite{Demkowicz14PRL,Zhou2021PRXQ} & $\Theta(N)$ & $\Theta(N)$ \\
\bottomrule
\end{tabular}
\end{table}
Unlike the bit flip noise model, for the amplitude damping noise it was only known that attaining the SQL requires an unbounded memory cost linear in $N$ \cite{Zhou2021PRXQ}. Here with bounded ancilla, we identified several strategies yielding QFI following the linear trend for large $N$.

\begin{figure}[!htbp]
    \centering
    \includegraphics[width=0.48\textwidth]
    {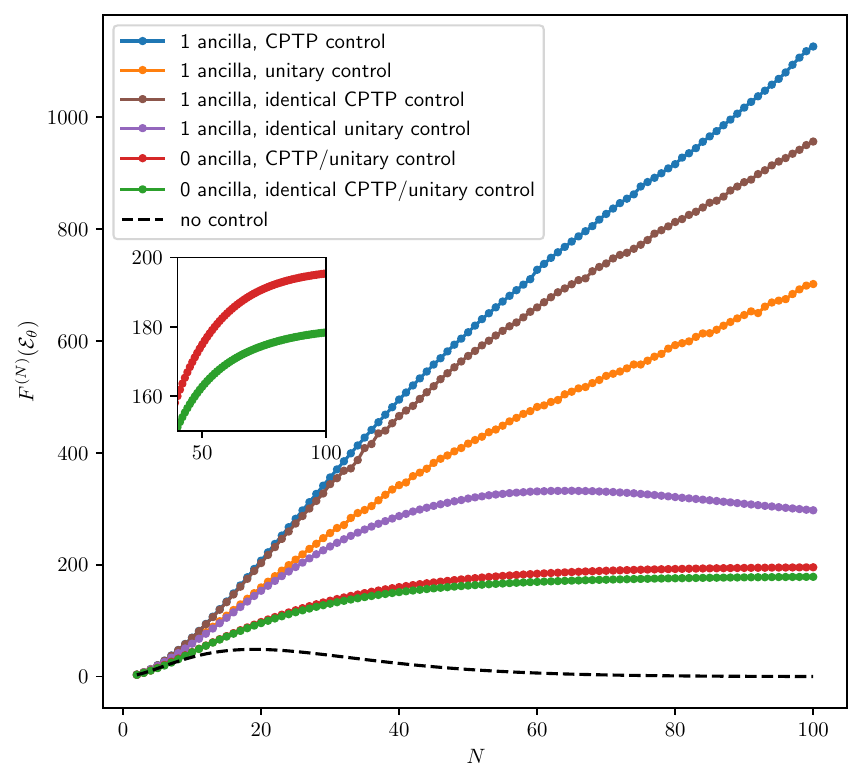}
    \caption{The QFI $F^{(N)}(\mathcal E_\theta)$ versus the number of queries $N$ to $\mathcal E_\theta$ for phase estimation with the amplitude damping noise. We zoom in on the interval $[40,100]$ of $N$ to more clearly demonstrate the advantage of using different control over identical control with ancilla-free strategies. As a benchmark, the dashed black line is obtained by an ancilla-free and control-free strategy with the input state $\ket{{+}}=\frac{1}{\sqrt{2}}(\ket{0} + \ket{1})$.}
    \label{fig:phase estimation AD comparison}
\end{figure}

We take $p=0.1$ and the true value of the parameter $\theta_0=1.0$, and optimize the QFI $F^{(N)}(\mathcal E_\theta)$ with different strategies, as illustrated in Fig.~\ref{fig:phase estimation AD comparison}. Since there is no known strategy with bounded ancilla for this noise model, we use as a benchmark the QFI obtained by a control-free strategy with the input state $\ket{+}=\frac{1}{\sqrt{2}}(\ket{0} + \ket{1})$, represented as a dashed line. In the ancilla-free case, we note that using different control operations can provide a slight enhancement compared to identical control operations, but we find no performance gaps between CPTP and unitary control. Assisted by one ancilla qubit, using CPTP control yields a QFI $F^{(N)}(\mathcal E_\theta)$ nearly linear in $N$ when $N$ is large. It may be interesting to theoretically investigate whether the SQL is achievable with one ancilla, which is an open problem so far. Moreover, unitary control in the ancilla-assisted case can significantly outperform generic CPTP control in the ancilla-free case. We also remark that all the $4$ control strategies have different performances in the ancilla-assisted case.

\subsection{Dephasing direction estimation}
In preceding examples we have applied our approach to phase estimation in the presence of noise. Here we consider a problem of non-unitary encoding, where we would like to estimate the dephasing direction angle $\theta$ from $N$ queries to $\mathcal E_\theta$ characterized by the Kraus operators
\begin{multline}
    K^{\theta,(\mathrm{dep})}_1 = \sqrt{1-p}I,\\ K^{\theta,(\mathrm{dep})}_2 =\sqrt{p}\left(\cos\theta \sigma_z + \sin\theta \sigma_x\right),
\end{multline}
for $0 \le p \le 1$. Known scalings of QFI for this model is the same as the scalings presented in Table \ref{tab:QFI scalings BF}, implying that the HL is attainable by QEC.

\begin{figure}[!htbp]
    \centering
    \includegraphics[width=0.48\textwidth]{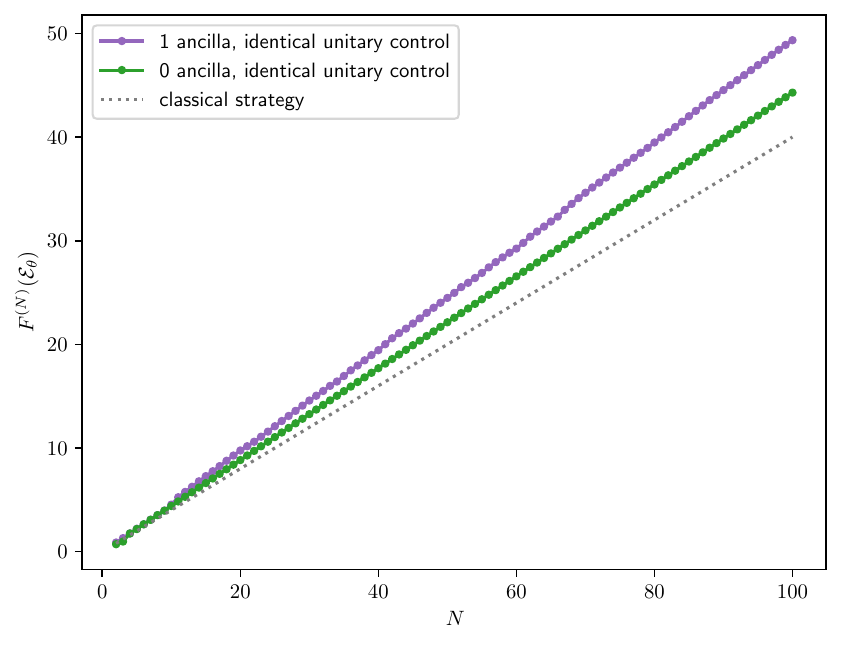}
    \caption{The QFI $F^{(N)}(\mathcal E_\theta)$ versus the number of queries $N$ to $\mathcal E_\theta$ for dephasing direction angle estimation. The classical strategy (dotted gray line) independently estimates $N$ channels, yielding a total QFI as the $N$-fold single channel QFI $NF^{(1)}(\mathcal E_\theta)$.}
    \label{fig:dephasing direction angle estimation}
\end{figure}

We take $p=0.1$ and the ground truth $\theta_0=1.0$, and apply our algorithm to optimize the QFI $F^{(N)}(\mathcal E_\theta)$ with restricted ancilla and control. Specifically, here we focus on the QFI attainable by identical unitary control with zero or one ancilla qubit, as more complicated control strategies do not offer large advantages. The results of QFI versus $N$ are plotted in Fig.~\ref{fig:dephasing direction angle estimation}. Clearly, there is an advantage of employing coherence, 
as using identical unitary control can outperform a classical strategy where $N$ channels are estimated independently, and using ancilla is indeed advantageous.

\section{Discussion}
We have provided an efficient approach to optimizing control-enhanced metrological strategies by leveraging tensor networks. With its high flexibility for applying arbitrary or identical control operations and assigning the ancilla dimension, we anticipate that this approach is well-suited for real-world experimental demonstration under certain restrictions. This may be particularly suitable for optical experiments, where typically the same control is applied each time with no or small ancilla in an optical loop \cite{Hou2019PRL,Hou2021PRL}.

Our approach naturally extends its applicability beyond single-parameter quantum channel estimation. An immediate application is in Hamiltonian parameter estimation with Markovian noise \cite{Demkowicz-Dobrza2017PRX,Zhou2018NC,Sekatski2017quantummetrology}, where our framework can be used to identify how the frequency of fast quantum control affects the metrological performance. We can also relax the assumption of estimating $N$ independent channels, but investigate the case where the non-Markovian noise takes effect, induced by the environment which has memory effect \cite{Chin2012PRL,Yang2019PRL,Altherr21PRL}. In this case the non-Markovian process can still be expressed as an MPO, with the bond dimension corresponding to the memory \cite{Pollock2018PRA,Pollock18PRL,Butler24PRL}, so our formalism can also be applicable. Besides, it could be interesting to generalize our framework to multi-parameter Hamiltonian estimation, where the asymptotically optimal QEC protocol has been identified by SDP previously \cite{Gorecki2020optimalprobeserror}. Finally, an exciting future work could be developing an efficient approach, possibly based on tensor networks combining the techniques of Ref.~\cite{Chabuda2020NC} and our work, to optimizing quantum metrology for many-body systems with multi-step control, which handles both spatial complexity and temporal complexity in a unified framework.

\section*{Acknowledgements}
This work is supported 
by the National Natural Science Foundation of China via Excellent Young Scientists Fund (Hong Kong and Macau) Project 12322516, Guangdong Basic and Applied Basic Research Foundation (Project No.~2022A1515010340), Guangdong Provincial Quantum Science Strategic Initiative (No.~GDZX2303007), and the Hong Kong Research Grant Council (RGC) through the Early Career Scheme (ECS) grant 27310822 and the General Research Fund (GRF) grant 17303923. The numerical results are obtained via the Python packages $\texttt{opt\_einsum}$ \cite{Smith2018opteinsum}, CVXPY \cite{diamond2016cvxpy,agrawal2018rewriting} with the solver MOSEK \cite{mosek}, and PennyLane \cite{bergholm2022pennylane}. The code accompanying the paper is available on GitHub \cite{code_note}.

\emph{Note added}.---After the completion of this work, we learned about an independent work using similar approaches \cite{kurdzialek2024quantum}.

\bibliographystyle{quantum}

\begin{thebibliography}{10}

\bibitem{giovannetti2004quantum}
Vittorio Giovannetti, Seth Lloyd, and Lorenzo Maccone.
\newblock ``Quantum-enhanced measurements: Beating the standard quantum limit''.
\newblock \href{https://dx.doi.org/10.1126/science.1104149}{Science {\bf 306}, 1330--1336}~(2004).

\bibitem{Giovannetti2006PRL}
Vittorio Giovannetti, Seth Lloyd, and Lorenzo Maccone.
\newblock ``{Quantum Metrology}''.
\newblock \href{https://dx.doi.org/10.1103/PhysRevLett.96.010401}{Phys. Rev. Lett. {\bf 96}, 010401}~(2006).

\bibitem{Degen17RMP}
C.~L. Degen, F.~Reinhard, and P.~Cappellaro.
\newblock ``Quantum sensing''.
\newblock \href{https://dx.doi.org/10.1103/RevModPhys.89.035002}{Rev. Mod. Phys. {\bf 89}, 035002}~(2017).

\bibitem{Huang23PRL}
Hsin-Yuan Huang, Yu~Tong, Di~Fang, and Yuan Su.
\newblock ``Learning many-body hamiltonians with heisenberg-limited scaling''.
\newblock \href{https://dx.doi.org/10.1103/PhysRevLett.130.200403}{Phys. Rev. Lett. {\bf 130}, 200403}~(2023).

\bibitem{Chu23PRLstrong}
Yaoming Chu, Xiangbei Li, and Jianming Cai.
\newblock ``Strong quantum metrological limit from many-body physics''.
\newblock \href{https://dx.doi.org/10.1103/PhysRevLett.130.170801}{Phys. Rev. Lett. {\bf 130}, 170801}~(2023).

\bibitem{Chabuda2020NC}
Krzysztof Chabuda, Jacek Dziarmaga, Tobias~J. Osborne, and Rafa{\l} Demkowicz-Dobrza{\'{n}}ski.
\newblock ``Tensor-network approach for quantum metrology in many-body quantum systems''.
\newblock \href{https://dx.doi.org/10.1038/s41467-019-13735-9}{Nat. Commun. {\bf 11}, 250}~(2020).

\bibitem{Yin2024heisenberglimited}
Chao Yin and Andrew Lucas.
\newblock ``Heisenberg-limited metrology with perturbing interactions''.
\newblock \href{https://dx.doi.org/10.22331/q-2024-03-28-1303}{{Quantum} {\bf 8}, 1303}~(2024).

\bibitem{Dutkiewicz2024advantageofquantum}
Alicja Dutkiewicz, Thomas~E. O'Brien, and Thomas Schuster.
\newblock ``The advantage of quantum control in many-body {H}amiltonian learning''.
\newblock \href{https://dx.doi.org/10.22331/q-2024-11-26-1537}{{Quantum} {\bf 8}, 1537}~(2024).

\bibitem{Fujiwara2008}
Akio Fujiwara and Hiroshi Imai.
\newblock ``A fibre bundle over manifolds of quantum channels and its application to quantum statistics''.
\newblock \href{https://dx.doi.org/10.1088/1751-8113/41/25/255304}{J. Phys. A {\bf 41}, 255304}~(2008).

\bibitem{Escher2011}
B.~M. Escher, R.~L. de~Matos~Filho, and L.~Davidovich.
\newblock ``General framework for estimating the ultimate precision limit in noisy quantum-enhanced metrology''.
\newblock \href{https://dx.doi.org/10.1038/nphys1958}{Nat. Phys. {\bf 7}, 406--411}~(2011).

\bibitem{Demkowicz-Dobrzanski2012}
Rafa{\l} Demkowicz-Dobrza{\'{n}}ski, Jan Ko{\l}ody{\'{n}}ski, and M{\u{a}}d{\u{a}}lin Gu{\c{t}}{\u{a}}.
\newblock ``The elusive heisenberg limit in quantum-enhanced metrology''.
\newblock \href{https://dx.doi.org/10.1038/ncomms2067}{Nat. Commun. {\bf 3}, 1063}~(2012).

\bibitem{Kolodynski_2013NJP}
Jan Kołodyński and Rafał Demkowicz-Dobrzański.
\newblock ``Efficient tools for quantum metrology with uncorrelated noise''.
\newblock \href{https://dx.doi.org/10.1088/1367-2630/15/7/073043}{New J. Phys. {\bf 15}, 073043}~(2013).

\bibitem{Demkowicz14PRL}
Rafal Demkowicz-Dobrza\ifmmode~\acute{n}\else \'{n}\fi{}ski and Lorenzo Maccone.
\newblock ``Using entanglement against noise in quantum metrology''.
\newblock \href{https://dx.doi.org/10.1103/PhysRevLett.113.250801}{Phys. Rev. Lett. {\bf 113}, 250801}~(2014).

\bibitem{Zhou2021PRXQ}
Sisi Zhou and Liang Jiang.
\newblock ``Asymptotic theory of quantum channel estimation''.
\newblock \href{https://dx.doi.org/10.1103/PRXQuantum.2.010343}{PRX Quantum {\bf 2}, 010343}~(2021).

\bibitem{Kurdzialek23PRL}
Stanis\l{}aw Kurdzia\l{}ek, Wojciech G\'orecki, Francesco Albarelli, and Rafa\l{} Demkowicz-Dobrza\ifmmode~\acute{n}\else \'{n}\fi{}ski.
\newblock ``Using adaptiveness and causal superpositions against noise in quantum metrology''.
\newblock \href{https://dx.doi.org/10.1103/PhysRevLett.131.090801}{Phys. Rev. Lett. {\bf 131}, 090801}~(2023).

\bibitem{Demkowicz-Dobrza2017PRX}
Rafa\l{} Demkowicz-Dobrza\ifmmode~\acute{n}\else \'{n}\fi{}ski, Jan Czajkowski, and Pavel Sekatski.
\newblock ``{Adaptive Quantum Metrology under General Markovian Noise}''.
\newblock \href{https://dx.doi.org/10.1103/PhysRevX.7.041009}{Phys. Rev. X {\bf 7}, 041009}~(2017).

\bibitem{Zhou2018NC}
Sisi Zhou, Mengzhen Zhang, John Preskill, and Liang Jiang.
\newblock ``Achieving the heisenberg limit in quantum metrology using quantum error correction''.
\newblock \href{https://dx.doi.org/10.1038/s41467-017-02510-3}{Nat. Commun. {\bf 9}, 78}~(2018).

\bibitem{Altherr21PRL}
Anian Altherr and Yuxiang Yang.
\newblock ``Quantum metrology for non-markovian processes''.
\newblock \href{https://dx.doi.org/10.1103/PhysRevLett.127.060501}{Phys. Rev. Lett. {\bf 127}, 060501}~(2021).

\bibitem{Liu23PRL}
Qiushi Liu, Zihao Hu, Haidong Yuan, and Yuxiang Yang.
\newblock ``Optimal strategies of quantum metrology with a strict hierarchy''.
\newblock \href{https://dx.doi.org/10.1103/PhysRevLett.130.070803}{Phys. Rev. Lett. {\bf 130}, 070803}~(2023).

\bibitem{Hou2019PRL}
Zhibo Hou, Rui-Jia Wang, Jun-Feng Tang, Haidong Yuan, Guo-Yong Xiang, Chuan-Feng Li, and Guang-Can Guo.
\newblock ``{Control-Enhanced Sequential Scheme for General Quantum Parameter Estimation at the Heisenberg Limit}''.
\newblock \href{https://dx.doi.org/10.1103/PhysRevLett.123.040501}{Phys. Rev. Lett. {\bf 123}, 040501}~(2019).

\bibitem{Hou2021PRL}
Zhibo Hou, Yan Jin, Hongzhen Chen, Jun-Feng Tang, Chang-Jiang Huang, Haidong Yuan, Guo-Yong Xiang, Chuan-Feng Li, and Guang-Can Guo.
\newblock ````super-heisenberg'' and heisenberg scalings achieved simultaneously in the estimation of a rotating field''.
\newblock \href{https://dx.doi.org/10.1103/PhysRevLett.126.070503}{Phys. Rev. Lett. {\bf 126}, 070503}~(2021).

\bibitem{ORUSAnnals}
Román Orús.
\newblock ``A practical introduction to tensor networks: Matrix product states and projected entangled pair states''.
\newblock \href{https://dx.doi.org/https://doi.org/10.1016/j.aop.2014.06.013}{Ann. Phys. (N. Y.) {\bf 349}, 117--158}~(2014).

\bibitem{Bridgeman_2017}
Jacob~C Bridgeman and Christopher~T Chubb.
\newblock ``Hand-waving and interpretive dance: an introductory course on tensor networks''.
\newblock \href{https://dx.doi.org/10.1088/1751-8121/aa6dc3}{J. Phys. A {\bf 50}, 223001}~(2017).

\bibitem{Torlai2023NC}
Giacomo Torlai, Christopher~J. Wood, Atithi Acharya, Giuseppe Carleo, Juan Carrasquilla, and Leandro Aolita.
\newblock ``Quantum process tomography with unsupervised learning and tensor networks''.
\newblock \href{https://dx.doi.org/10.1038/s41467-023-38332-9}{Nat. Commun. {\bf 14}, 2858}~(2023).

\bibitem{Guo20PRA}
Chu Guo, Kavan Modi, and Dario Poletti.
\newblock ``Tensor-network-based machine learning of non-markovian quantum processes''.
\newblock \href{https://dx.doi.org/10.1103/PhysRevA.102.062414}{Phys. Rev. A {\bf 102}, 062414}~(2020).

\bibitem{helstrom1976quantum}
Carl~W. Helstrom.
\newblock ``Quantum detection and estimation theory''.
\newblock Academic Press. New York~(1976).

\bibitem{holevo2011probabilistic}
Alexander~S. Holevo.
\newblock ``Probabilistic and statistical aspects of quantum theory''.
\newblock North-Holland Publishing Company. Amsterdam~(1982).

\bibitem{Braunstein1994PRL}
Samuel~L. Braunstein and Carlton~M. Caves.
\newblock ``{Statistical distance and the geometry of quantum states}''.
\newblock \href{https://dx.doi.org/10.1103/PhysRevLett.72.3439}{Phys. Rev. Lett. {\bf 72}, 3439--3443}~(1994).

\bibitem{macieszczak2013quantum}
Katarzyna Macieszczak.
\newblock ``Quantum fisher information: Variational principle and simple iterative algorithm for its efficient computation''~(2013).
\newblock  \href{http://arxiv.org/abs/1312.1356}{arXiv:1312.1356}.

\bibitem{Macieszczak_2014NJP}
Katarzyna Macieszczak, Martin Fraas, and Rafał Demkowicz-Dobrzański.
\newblock ``Bayesian quantum frequency estimation in presence of collective dephasing''.
\newblock \href{https://dx.doi.org/10.1088/1367-2630/16/11/113002}{New J. Phys. {\bf 16}, 113002}~(2014).

\bibitem{Len2022NC}
Yink~Loong Len, Tuvia Gefen, Alex Retzker, and Jan Ko{\l}ody{\'{n}}ski.
\newblock ``Quantum metrology with imperfect measurements''.
\newblock \href{https://dx.doi.org/10.1038/s41467-022-33563-8}{Nat. Commun. {\bf 13}, 6971}~(2022).

\bibitem{gutoski2007toward}
Gus Gutoski and John Watrous.
\newblock ``Toward a general theory of quantum games''.
\newblock In Proceedings of the Thirty-Ninth Annual ACM Symposium on Theory of Computing.
\newblock \href{https://dx.doi.org/10.1145/1250790.1250873}{Page 565–574}.
\newblock New York, NY, USA~(2007). Association for Computing Machinery.

\bibitem{Chiribella2008PRL}
G.~Chiribella, G.~M. D'Ariano, and P.~Perinotti.
\newblock ``Quantum {C}ircuit {A}rchitecture''.
\newblock \href{https://dx.doi.org/10.1103/PhysRevLett.101.060401}{Phys. Rev. Lett. {\bf 101}, 060401}~(2008).

\bibitem{Chiribella2009PRA}
Giulio Chiribella, Giacomo~Mauro D'Ariano, and Paolo Perinotti.
\newblock ``Theoretical framework for quantum networks''.
\newblock \href{https://dx.doi.org/10.1103/PhysRevA.80.022339}{Phys. Rev. A {\bf 80}, 022339}~(2009).

\bibitem{choi1975completely}
Man-Duen Choi.
\newblock ``Completely positive linear maps on complex matrices''.
\newblock Linear Algebra Appl. {\bf 10}, 285--290~(1975).

\bibitem{Crosswhite08PRA}
Gregory~M. Crosswhite and Dave Bacon.
\newblock ``Finite automata for caching in matrix product algorithms''.
\newblock \href{https://dx.doi.org/10.1103/PhysRevA.78.012356}{Phys. Rev. A {\bf 78}, 012356}~(2008).

\bibitem{mcculloch2008infinite}
I.~P. McCulloch.
\newblock ``Infinite size density matrix renormalization group, revisited''~(2008).
\newblock  \href{http://arxiv.org/abs/0804.2509}{arXiv:0804.2509}.

\bibitem{Hubig17PRB}
C.~Hubig, I.~P. McCulloch, and U.~Schollw\"ock.
\newblock ``Generic construction of efficient matrix product operators''.
\newblock \href{https://dx.doi.org/10.1103/PhysRevB.95.035129}{Phys. Rev. B {\bf 95}, 035129}~(2017).

\bibitem{WhiteDMRG92PRL}
Steven~R. White.
\newblock ``Density matrix formulation for quantum renormalization groups''.
\newblock \href{https://dx.doi.org/10.1103/PhysRevLett.69.2863}{Phys. Rev. Lett. {\bf 69}, 2863--2866}~(1992).

\bibitem{WhiteDMRG93PRB}
Steven~R. White.
\newblock ``Density-matrix algorithms for quantum renormalization groups''.
\newblock \href{https://dx.doi.org/10.1103/PhysRevB.48.10345}{Phys. Rev. B {\bf 48}, 10345--10356}~(1993).

\bibitem{Horodecki03EntanglementBreaking}
Michael Horodecki, Peter~W. Shor, and Mary~Beth Ruskai.
\newblock ``Entanglement breaking channels''.
\newblock \href{https://dx.doi.org/10.1142/S0129055X03001709}{Rev. Math. Phys. {\bf 15}, 629--641}~(2003).

\bibitem{Peres96PRL}
Asher Peres.
\newblock ``Separability criterion for density matrices''.
\newblock \href{https://dx.doi.org/10.1103/PhysRevLett.77.1413}{Phys. Rev. Lett. {\bf 77}, 1413--1415}~(1996).

\bibitem{HORODECKI19961}
Michał Horodecki, Paweł Horodecki, and Ryszard Horodecki.
\newblock ``Separability of mixed states: necessary and sufficient conditions''.
\newblock \href{https://dx.doi.org/https://doi.org/10.1016/S0375-9601(96)00706-2}{Phys. Lett. A {\bf 223}, 1--8}~(1996).

\bibitem{Gurvits03Classical}
Leonid Gurvits.
\newblock ``Classical deterministic complexity of edmonds' problem and quantum entanglement''.
\newblock In Proceedings of the Thirty-Fifth Annual ACM Symposium on Theory of Computing.
\newblock \href{https://dx.doi.org/10.1145/780542.780545}{Page 10–19}.
\newblock New York, NY, USA~(2003). Association for Computing Machinery.

\bibitem{Doherty02PRL}
A.~C. Doherty, Pablo~A. Parrilo, and Federico~M. Spedalieri.
\newblock ``Distinguishing separable and entangled states''.
\newblock \href{https://dx.doi.org/10.1103/PhysRevLett.88.187904}{Phys. Rev. Lett. {\bf 88}, 187904}~(2002).

\bibitem{Doherty04PRA}
Andrew~C. Doherty, Pablo~A. Parrilo, and Federico~M. Spedalieri.
\newblock ``Complete family of separability criteria''.
\newblock \href{https://dx.doi.org/10.1103/PhysRevA.69.022308}{Phys. Rev. A {\bf 69}, 022308}~(2004).

\bibitem{Yu2021NC}
Xiao-Dong Yu, Timo Simnacher, Nikolai Wyderka, H.~Chau Nguyen, and Otfried G{\"u}hne.
\newblock ``A complete hierarchy for the pure state marginal problem in quantum mechanics''.
\newblock \href{https://dx.doi.org/10.1038/s41467-020-20799-5}{Nat. Commun. {\bf 12}, 1012}~(2021).

\bibitem{Yu2022PRXQuantum}
Xiao-Dong Yu, Timo Simnacher, H.~Chau Nguyen, and Otfried G\"uhne.
\newblock ``Quantum-inspired hierarchy for rank-constrained optimization''.
\newblock \href{https://dx.doi.org/10.1103/PRXQuantum.3.010340}{PRX Quantum {\bf 3}, 010340}~(2022).

\bibitem{Cincio13PRL}
L.~Cincio and G.~Vidal.
\newblock ``Characterizing topological order by studying the ground states on an infinite cylinder''.
\newblock \href{https://dx.doi.org/10.1103/PhysRevLett.110.067208}{Phys. Rev. Lett. {\bf 110}, 067208}~(2013).

\bibitem{Corboz16PRB}
Philippe Corboz.
\newblock ``Variational optimization with infinite projected entangled-pair states''.
\newblock \href{https://dx.doi.org/10.1103/PhysRevB.94.035133}{Phys. Rev. B {\bf 94}, 035133}~(2016).

\bibitem{Forsythe1977computer}
George~E. Forsythe, Michael~A. Malcolm, and Cleve~B. Moler.
\newblock ``Computer methods for mathematical computations''.
\newblock Prentice Hall Professional Technical Reference. ~(1977).

\bibitem{Brent2013algorithms}
Richard~P. Brent.
\newblock ``Algorithms for minimization without derivatives''.
\newblock Courier Corporation. ~(2013).

\bibitem{2020SciPy-NMeth}
Pauli~Virtanen \textit{et al.}
\newblock ``{{SciPy} 1.0: Fundamental Algorithms for Scientific Computing in Python}''.
\newblock \href{https://dx.doi.org/10.1038/s41592-019-0686-2}{Nat. Methods {\bf 17}, 261--272}~(2020).

\bibitem{Koczor_2020}
Bálint Koczor, Suguru Endo, Tyson Jones, Yuichiro Matsuzaki, and Simon~C Benjamin.
\newblock ``Variational-state quantum metrology''.
\newblock \href{https://dx.doi.org/10.1088/1367-2630/ab965e}{New J. Phys. {\bf 22}, 083038}~(2020).

\bibitem{Yang2020}
Xiaodong Yang, Jayne Thompson, Ze~Wu, Mile Gu, Xinhua Peng, and Jiangfeng Du.
\newblock ``Probe optimization for quantum metrology via closed-loop learning control''.
\newblock \href{https://dx.doi.org/10.1038/s41534-020-00292-z}{npj Quantum Inf. {\bf 6}, 62}~(2020).

\bibitem{Ma21IEEE}
Ziqi Ma, Pranav Gokhale, Tian-Xing Zheng, Sisi Zhou, Xiaofei Yu, Liang Jiang, Peter Maurer, and Frederic~T. Chong.
\newblock ``Adaptive circuit learning for quantum metrology''.
\newblock In 2021 IEEE International Conference on Quantum Computing and Engineering (QCE).
\newblock \href{https://dx.doi.org/10.1109/QCE52317.2021.00063}{Pages 419--430}.
\newblock ~(2021).

\bibitem{Meyer2021}
Johannes~Jakob Meyer, Johannes Borregaard, and Jens Eisert.
\newblock ``A variational toolbox for quantum multi-parameter estimation''.
\newblock \href{https://dx.doi.org/10.1038/s41534-021-00425-y}{npj Quantum Inf. {\bf 7}, 89}~(2021).

\bibitem{Beckey22PRRvariational}
Jacob~L. Beckey, M.~Cerezo, Akira Sone, and Patrick~J. Coles.
\newblock ``Variational quantum algorithm for estimating the quantum fisher information''.
\newblock \href{https://dx.doi.org/10.1103/PhysRevResearch.4.013083}{Phys. Rev. Res. {\bf 4}, 013083}~(2022).

\bibitem{Le2023variational}
Trung~Kien Le, Hung~Q. Nguyen, and Le~Bin Ho.
\newblock ``Variational quantum metrology for multiparameter estimation under dephasing noise''.
\newblock \href{https://dx.doi.org/10.1038/s41598-023-44786-0}{Sci. Rep. {\bf 13}, 17775}~(2023).

\bibitem{Dominik17PRA}
Dominik \ifmmode~\check{S}\else \v{S}\fi{}afr\'anek.
\newblock ``Discontinuities of the quantum fisher information and the bures metric''.
\newblock \href{https://dx.doi.org/10.1103/PhysRevA.95.052320}{Phys. Rev. A {\bf 95}, 052320}~(2017).

\bibitem{Seveso_2020}
Luigi Seveso, Francesco Albarelli, Marco~G Genoni, and Matteo G~A Paris.
\newblock ``On the discontinuity of the quantum fisher information for quantum statistical models with parameter dependent rank''.
\newblock \href{https://dx.doi.org/10.1088/1751-8121/ab599b}{J. Phys. A {\bf 53}, 02LT01}~(2019).

\bibitem{zhou2019exact}
Sisi Zhou and Liang Jiang.
\newblock ``An exact correspondence between the quantum fisher information and the bures metric''~(2019).
\newblock  \href{http://arxiv.org/abs/1910.08473}{arXiv:1910.08473}.

\bibitem{Ye22PRA}
Yating Ye and Xiao-Ming Lu.
\newblock ``Quantum cram\'er-rao bound for quantum statistical models with parameter-dependent rank''.
\newblock \href{https://dx.doi.org/10.1103/PhysRevA.106.022429}{Phys. Rev. A {\bf 106}, 022429}~(2022).

\bibitem{zhou2024limits}
Sisi Zhou.
\newblock ``Limits of noisy quantum metrology with restricted quantum controls''.
\newblock \href{https://dx.doi.org/10.1103/PhysRevLett.133.170801}{Phys. Rev. Lett. {\bf 133}, 170801}~(2024).

\bibitem{Kessler14PRL}
E.~M. Kessler, I.~Lovchinsky, A.~O. Sushkov, and M.~D. Lukin.
\newblock ``Quantum error correction for metrology''.
\newblock \href{https://dx.doi.org/10.1103/PhysRevLett.112.150802}{Phys. Rev. Lett. {\bf 112}, 150802}~(2014).

\bibitem{Sekatski2017quantummetrology}
Pavel Sekatski, Michalis Skotiniotis, Janek Ko{\l{}}ody{\'{n}}ski, and Wolfgang D{\"{u}}r.
\newblock ``Quantum metrology with full and fast quantum control''.
\newblock \href{https://dx.doi.org/10.22331/q-2017-09-06-27}{{Quantum} {\bf 1}, 27}~(2017).

\bibitem{Chin2012PRL}
Alex~W. Chin, Susana~F. Huelga, and Martin~B. Plenio.
\newblock ``Quantum metrology in non-markovian environments''.
\newblock \href{https://dx.doi.org/10.1103/PhysRevLett.109.233601}{Phys. Rev. Lett. {\bf 109}, 233601}~(2012).

\bibitem{Yang2019PRL}
Yuxiang Yang.
\newblock ``Memory effects in quantum metrology''.
\newblock \href{https://dx.doi.org/10.1103/PhysRevLett.123.110501}{Phys. Rev. Lett. {\bf 123}, 110501}~(2019).

\bibitem{Pollock2018PRA}
Felix~A. Pollock, C\'esar Rodr\'{\i}guez-Rosario, Thomas Frauenheim, Mauro Paternostro, and Kavan Modi.
\newblock ``Non-markovian quantum processes: Complete framework and efficient characterization''.
\newblock \href{https://dx.doi.org/10.1103/PhysRevA.97.012127}{Phys. Rev. A {\bf 97}, 012127}~(2018).

\bibitem{Pollock18PRL}
Felix~A. Pollock, C\'esar Rodr\'{\i}guez-Rosario, Thomas Frauenheim, Mauro Paternostro, and Kavan Modi.
\newblock ``Operational markov condition for quantum processes''.
\newblock \href{https://dx.doi.org/10.1103/PhysRevLett.120.040405}{Phys. Rev. Lett. {\bf 120}, 040405}~(2018).

\bibitem{Butler24PRL}
Eoin~P. Butler, Gerald~E. Fux, Carlos Ortega-Taberner, Brendon~W. Lovett, Jonathan Keeling, and Paul~R. Eastham.
\newblock ``Optimizing performance of quantum operations with non-markovian decoherence: The tortoise or the hare?''.
\newblock \href{https://dx.doi.org/10.1103/PhysRevLett.132.060401}{Phys. Rev. Lett. {\bf 132}, 060401}~(2024).

\bibitem{Gorecki2020optimalprobeserror}
Wojciech G{\'{o}}recki, Sisi Zhou, Liang Jiang, and Rafa{\l{}} Demkowicz-Dobrza{\'{n}}ski.
\newblock ``Optimal probes and error-correction schemes in multi-parameter quantum metrology''.
\newblock \href{https://dx.doi.org/10.22331/q-2020-07-02-288}{{Quantum} {\bf 4}, 288}~(2020).

\bibitem{Smith2018opteinsum}
Daniel~G. a.~Smith and Johnnie Gray.
\newblock ``opt\_einsum - a python package for optimizing contraction order for einsum-like expressions''.
\newblock \href{https://dx.doi.org/10.21105/joss.00753}{J. Open Source Softw. {\bf 3}, 753}~(2018).

\bibitem{diamond2016cvxpy}
Steven Diamond and Stephen Boyd.
\newblock ``Cvxpy: a python-embedded modeling language for convex optimization''.
\newblock J. Mach. Learn. Res. {\bf 17}, 2909–2913~(2016).
\newblock  url:~\url{https://dl.acm.org/doi/10.5555/2946645.3007036}.

\bibitem{agrawal2018rewriting}
Steven~Diamond Akshay~Agrawal, Robin~Verschueren and Stephen Boyd.
\newblock ``A rewriting system for convex optimization problems''.
\newblock \href{https://dx.doi.org/10.1080/23307706.2017.1397554}{J. Control Decis. {\bf 5}, 42--60}~(2018).

\bibitem{mosek}
{MOSEK ApS}.
\newblock ``The mosek optimizer api for python manual. version 9.3.''.
\newblock ~(2021).
\newblock  url:~\url{https://docs.mosek.com/9.3/pythonapi/index.html}.

\bibitem{bergholm2022pennylane}
Ville~Bergholm \textit{et al.}
\newblock ``Pennylane: Automatic differentiation of hybrid quantum-classical computations''~(2022).
\newblock  \href{http://arxiv.org/abs/1811.04968}{arXiv:1811.04968}.

\bibitem{code_note}
Qiushi Liu and Yuxiang Yang.
\newblock ``Tensor network algorithm for optimizing control-enhanced quantum metrology''.
\newblock \url{https://github.com/qiushi-liu/tensor_network_metrology}~(2024).

\bibitem{kurdzialek2024quantum}
Stanislaw Kurdzialek, Piotr Dulian, Joanna Majsak, Sagnik Chakraborty, and Rafal Demkowicz-Dobrzanski.
\newblock ``Quantum metrology using quantum combs and tensor network formalism''~(2024).
\newblock  \href{http://arxiv.org/abs/2403.04854}{arXiv:2403.04854}.

\bibitem{Nielsen_Chuang_2010}
Michael~A. Nielsen and Isaac~L. Chuang.
\newblock ``Quantum computation and quantum information: 10th anniversary edition''.
\newblock Cambridge University Press. Cambridge~(2010).

\bibitem{Duchi11JMLR}
John Duchi, Elad Hazan, and Yoram Singer.
\newblock ``Adaptive subgradient methods for online learning and stochastic optimization''.
\newblock J. Mach. Learn. Res. {\bf 12}, 2121–2159~(2011).
\newblock  url:~\url{https://dl.acm.org/doi/10.5555/1953048.2021068}.

\end{thebibliography}
\providecommand{\noopsort}[1]{}\providecommand{\singleletter}[1]{#1}%

\onecolumn
\appendix

\section{Proof of Eq.~(\ref{eq:MPO derivative})} \label{app:MPO derivative}
It is not difficult to verify that Eq.~(\ref{eq:MPO derivative}) is equivalent to $dE_\theta^{\otimes N}/d\theta=\sum_{i=1}^N E_\theta^{\otimes i-1} \otimes \dot E_\theta \otimes E_\theta^{\otimes N-i}$. Nevertheless, here we provide a diagrammatic method \cite{Crosswhite08PRA} to derive this result in an intuitive way, which has been widely used in compressing sum of local Hamiltonian terms into an MPO.

\begin{figure}[!htbp]
    \centering
    \includegraphics[width=0.4\textwidth]{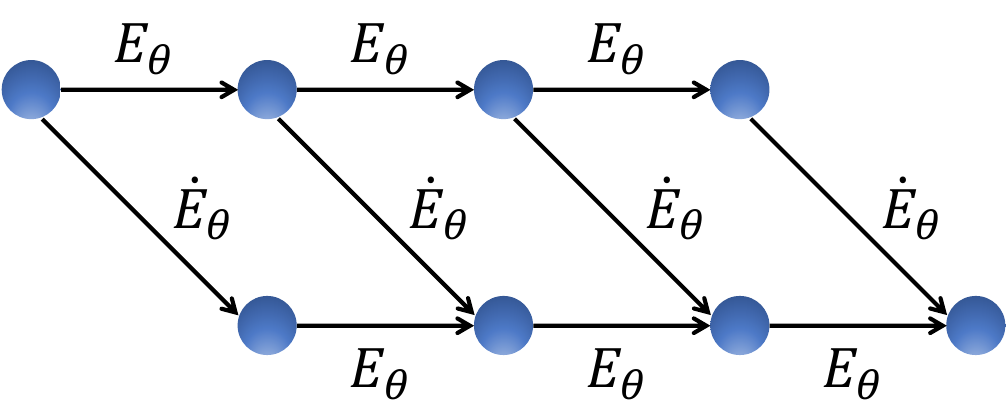}
    \caption{Matrix product diagram for expressing $dE_\theta^{\otimes 4}/d\theta$.}
    \label{fig:directed_graph_MPO}
\end{figure}

Without loss of generality, let us consider the case of $N=4$. Then $dE_\theta^{\otimes 4}/d\theta=\sum_{i=1}^4 E_\theta^{\otimes i-1} \otimes \dot E_\theta \otimes E_\theta^{\otimes 4-i}$ can be schematically illustrated as a directed graph in Fig.~\ref{fig:directed_graph_MPO}. Each term in in the summation $\sum_{i=1}^4 E_\theta^{\otimes i-1} \otimes \dot E_\theta \otimes E_\theta^{\otimes 4-i}$ is represented by a path from the top left vertex to the bottom right vertex. 

Now we see the diagram as a matrix product in this way: each vertex represents a possible value of an index, and $2$ vertices in the same column correspond to $2$ possible values of an index. A directed edge from vertex $i$ to $j$ represents a nonzero matrix element (weight) $M_{ij}$. The diagram can thus be interpreted as a product of $4$ matrices $M_1,M_2,M_3,M_4$ (our convention takes the reverse order):
\begin{equation}
    (M_4)_{1\gamma_3}(M_3)_{\gamma_3\gamma_2}(M_2)_{\gamma_2\gamma_1}(M_1)_{\gamma_11},
\end{equation}
where each index $\gamma_i$ can take two values $1$ and $2$. According to Fig.~\ref{fig:directed_graph_MPO}, we have
\begin{equation}
    M_4 = \begin{pmatrix}
        E_\theta & \dot E_\theta
    \end{pmatrix},\ M_3 =M_2= \begin{pmatrix}
        E_\theta & \dot E_\theta \\
        0 & E_\theta
    \end{pmatrix},\ M_1 = \begin{pmatrix}
        \dot E_\theta \\ E_\theta
    \end{pmatrix}.
\end{equation}
Noting that each $E_\theta$ or $\dot E_\theta$ is regarded as a tensor with $4$ indices (two $\alpha$ indices and two $\beta$ indices) in Eq.~(\ref{eq:MPO derivative}), we thus complete the proof for $N=4$. The same argument can be straightforwardly generalized to any $N$.

\section{Tensor networks for $f_1$ and $f_2$ with bounded ancilla} \label{app: f1 and f2 with ancilla}
Apparently, if the ancilla $A$ is accessible, one can replace $\mathcal E_\theta$ by $\mathcal E_\theta \otimes \mathcal I_A$ and apply the tensor contraction in Fig.~\ref{fig:tensor networks for f_1 and f_2}. Nevertheless, explicitly contracting the identity operator with other tensors is unnecessary and can be safely omitted. Instead, we can make a direct connection between indices related by an identity channel.

\begin{figure} [!htbp]
\centering
\subfigure[Tensor network for $f_2$.]{\includegraphics[width=0.6\textwidth]{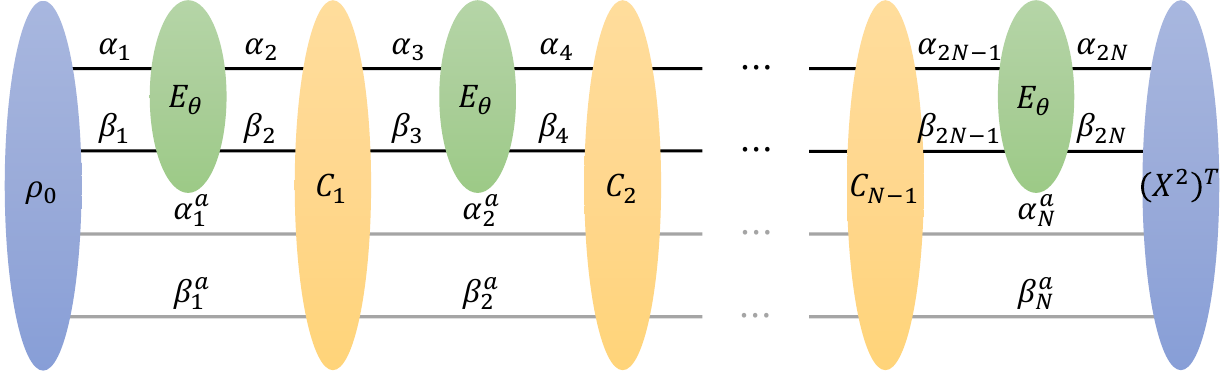}\label{fig:f_2 with ancilla}} \\
\subfigure[Tensor network for $f_1$.]{\includegraphics[width=0.6\textwidth]{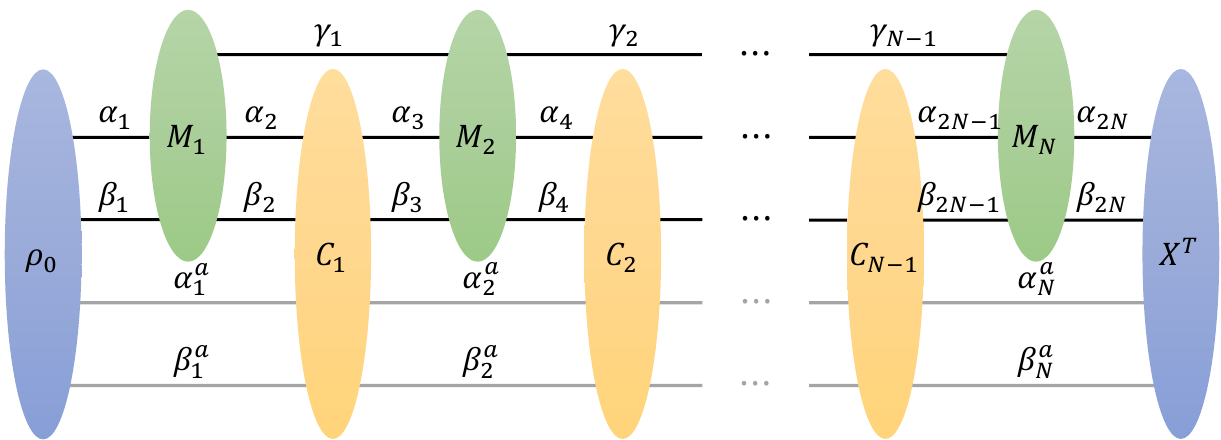}\label{fig:f_1 with ancilla}}
\caption{Tensor networks for computing $f_2$ and $f_1$ with bounded ancilla.}
\label{fig:tensor networks for f_1 and f_2 with ancilla}
\end{figure}

We take the input state $\rho_0=(\rho_0)^{\alpha_1\alpha_1^a}_{\beta_1\beta_1^a}\ketbra{\alpha_1\alpha_1^a}{\beta_1\beta_1^a}$ and the Choi operator of the $i$-th control operation $C_i=(C_i)^{\alpha_{2i+1}\alpha_{i+1}^a \alpha_{2i}\alpha_{i}^a}_{\beta_{2i+1}\beta_{i+1}^a \beta_{2i}\beta_{i}^a}\ketbra{\alpha_{2i+1}\alpha_{i+1}^a \alpha_{2i}\alpha_{i}^a}{\beta_{2i+1}\beta_{i+1}^a \beta_{2i}\beta_{i}^a}$, where the superscript ``$a$'' is associated with the ancilla and we have followed the Einstein summation convention. We then have
\begin{equation}
    f_2 = (X^2)^{\beta_{2N}\beta_{N}^a}_{\alpha_{2N}\alpha_{N}^a} (E_\theta)^{\alpha_{2N} \alpha_{2N-1}}_{\beta_{2N} \beta_{2N-1}} (C_{N-1})^{\alpha_{2N-1} \alpha_N^a \alpha_{2N-2} \alpha_{N-1}^a}_{\beta_{2N-1} \beta_N^a \beta_{2N-2} \beta_{N-1}^a} \cdots (C_1)^{\alpha_3 \alpha_2^a \alpha_2 \alpha_1^a}_{\beta_3 \beta_2^a \beta_2 \beta_1^a}(E_\theta)^{\alpha_2 \alpha_1}_{\beta_2 \beta_1}(\rho_0)^{\alpha_1\alpha_1^a}_{\beta_1\beta_1^a},
\end{equation}
illustrated in Fig.~\ref{fig:f_2 with ancilla} and 
\begin{multline}
    f_1 = X^{\beta_{2N}\beta_{N}^a}_{\alpha_{2N}\alpha_{N}^a} (M_N)^{\alpha_{2N} \alpha_{2N-1}}_{\beta_{2N} \beta_{2N-1} \gamma_{N-1}} (C_{N-1})^{\alpha_{2N-1} \alpha_N^a \alpha_{2N-2} \alpha_{N-1}^a}_{\beta_{2N-1} \beta_N^a \beta_{2N-2} \beta_{N-1}^a} \cdots \\
    (M_2)^{\alpha_{4} \alpha_{3}}_{\beta_{4} \beta_{3} \gamma_{2}\gamma_{1}} (C_1)^{\alpha_3 \alpha_2^a \alpha_2 \alpha_1^a}_{\beta_3 \beta_2^a \beta_2 \beta_1^a} (M_1)^{\alpha_{2} \alpha_{1}}_{\beta_{2} \beta_{1} \gamma_{1}} (\rho_0)^{\alpha_1\alpha_1^a}_{\beta_1\beta_1^a},
\end{multline}
illustrated in Fig.~\ref{fig:f_1 with ancilla}.

\section{Variational circuit ansatz} \label{app:variational ansatz}
In this work we employ a simple variational ansatz $U(\boldsymbol{\phi}_i)$ for the control operation $C_i$. For simplicity we assume that all the control operations share the same circuit layout (possibly with different parameters $\boldsymbol{\phi}_i$). Now we can omit the subscript and focus on one control operation $U(\boldsymbol{\phi})$.

\begin{figure} [!htbp]
\centering
\subfigure[Variational unitary.]{\includegraphics[height=0.15\textwidth]{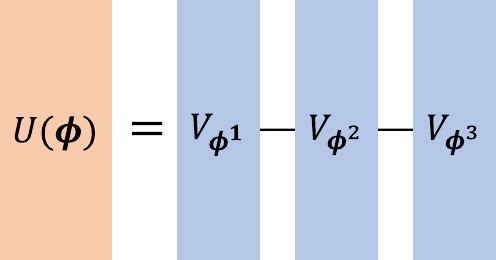}\label{fig:variational unitary}} \hspace{3em}
\subfigure[Variational layer.]{\includegraphics[height=0.15\textwidth]{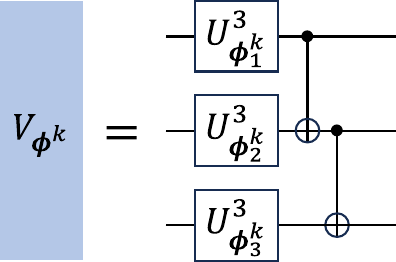}\label{fig:variational layer}}
\caption{Variational ansatz for $n=3$ qubits and $l=3$ layers.}
\label{fig:variational ansatz}
\end{figure}

As illustrated in Fig.~\ref{fig:variational ansatz}, our ansatz for the control $U(\boldsymbol{\phi})$ consists of $l$ layers $V_{\boldsymbol{\phi}^1},\dots,V_{\boldsymbol{\phi}^l}$, and each layer comprises $n$ single-qubit unitary gates $U^3$ (characterzed by $3$ rotational angles using the ZYZ decomposition \cite{Nielsen_Chuang_2010}) and $n-1$ CNOT gates connecting neighboring qubits.

In numerical experiments of estimating qubit channels, we take $n=1$, $l=1$ for the ancilla-free case and $n=2$, $l=3$ for the one-ancilla-assisted case. The variational parameters are optimized by the Adagrad algorithm \cite{Duchi11JMLR}.
\end{document}